\documentclass[prl,aps,preprintnumbers,superscriptaddress,nofootinbib,floatfix,twocolumn,notitlepage]{revtex4-1}

\usepackage{graphicx,epsfig,psfrag,amssymb,hyperref}
\usepackage{multirow}
\usepackage{color,graphicx,epsfig,psfrag,amsmath,empheq}
\usepackage{bm}
\usepackage{mathrsfs,amsfonts,, color}
\usepackage{slashed}
\usepackage[caption=false]{subfig}
\usepackage{hepunits}
\usepackage{color}

\usepackage{pdfpages}

\definecolor{darkgreen}{rgb}{0,0.5,0}
\definecolor{darkblue}{rgb}{0,0,0.5}
\definecolor{newred}{rgb}{0.5,0.1,0}
\definecolor{gold}{rgb}{0.7,0.7,0}

\newcommand{\Ref}[1]{\cite{#1}}
\newcommand{\Refs}[1]{\cite{#1}}

\newcommand{\Fig}[1]{Fig.~\ref{#1}}

\newcommand{\be}{\begin{equation}}
\newcommand{\ee}{\end{equation}}
\def\lsim{\mathrel{\rlap{\lower4pt\hbox{\hskip 0.5 pt$\sim$}}
\raise1pt\hbox{$<$}}}

\newcommand{\cO}{\mathcal{O}}
\newcommand{\cL}{\mathcal{L}}

\newcommand{\cR}{\mathcal{R}}

\newcommand{\mev}{\ensuremath{\mathrm{\: Me\kern -0.1em V}}\xspace}
\newcommand{\gev}{\ensuremath{\mathrm{\: Ge\kern -0.1em V}}\xspace}
\newcommand{\tev}{\ensuremath{\mathrm{\: Te\kern -0.1em V}}\xspace}

\begin{document}

\title{Inclusive Dark Photon Search at LHCb}

\author{Philip Ilten}
\email{philten@cern.ch}
\affiliation{Laboratory for Nuclear Science, Massachusetts Institute of Technology, Cambridge, MA 02139, U.S.A.}

\author{Yotam Soreq}
\email{soreqy@mit.edu}
\affiliation{Center for Theoretical Physics, Massachusetts Institute of Technology, Cambridge, MA 02139, U.S.A.}

\author{Jesse Thaler}
\email{jthaler@mit.edu}
\affiliation{Center for Theoretical Physics, Massachusetts Institute of Technology, Cambridge, MA 02139, U.S.A.}

\author{Mike Williams}
\email{mwill@mit.edu}
\affiliation{Laboratory for Nuclear Science, Massachusetts Institute of Technology, Cambridge, MA 02139, U.S.A.}

\author{Wei Xue}
\email{weixue@mit.edu}
\affiliation{Center for Theoretical Physics, Massachusetts Institute of Technology, Cambridge, MA 02139, U.S.A.}

\begin{abstract}
We propose an inclusive search for dark photons $A'$ at the LHCb experiment based on both prompt and displaced di-muon resonances.
Because the couplings of the dark photon are inherited from the photon via kinetic mixing, the dark photon $A' \to \mu^+ \mu^-$ rate can be directly inferred from the off-shell photon $\gamma^* \to \mu^+ \mu^-$ rate, making this a fully data-driven search.
For Run~3 of the LHC, we estimate that LHCb will have sensitivity to large regions of the unexplored dark-photon parameter space, especially in the $210$--$520\mev$ and $10$--$40\gev$ mass ranges.
This search leverages the excellent invariant-mass and vertex resolution of LHCb, along with its unique particle-identification and real-time data-analysis capabilities.
\end{abstract}

\preprint{MIT-CTP 4785}
\maketitle

Dark matter---firmly established through its interactions with gravity---remains an enigma.
Though there are increasingly stringent constraints on direct couplings between visible matter and dark matter, little is known about the dynamics within the dark sector itself.
An intriguing possibility is that dark matter might interact via a new dark force, felt only feebly by standard model (SM) particles.
This has motivated a worldwide effort to search for dark forces and other portals between the visible and dark sectors (see \Ref{Essig:2013lka} for a review).

\begin{figure*}[!t]
\includegraphics[width=1.7\columnwidth]{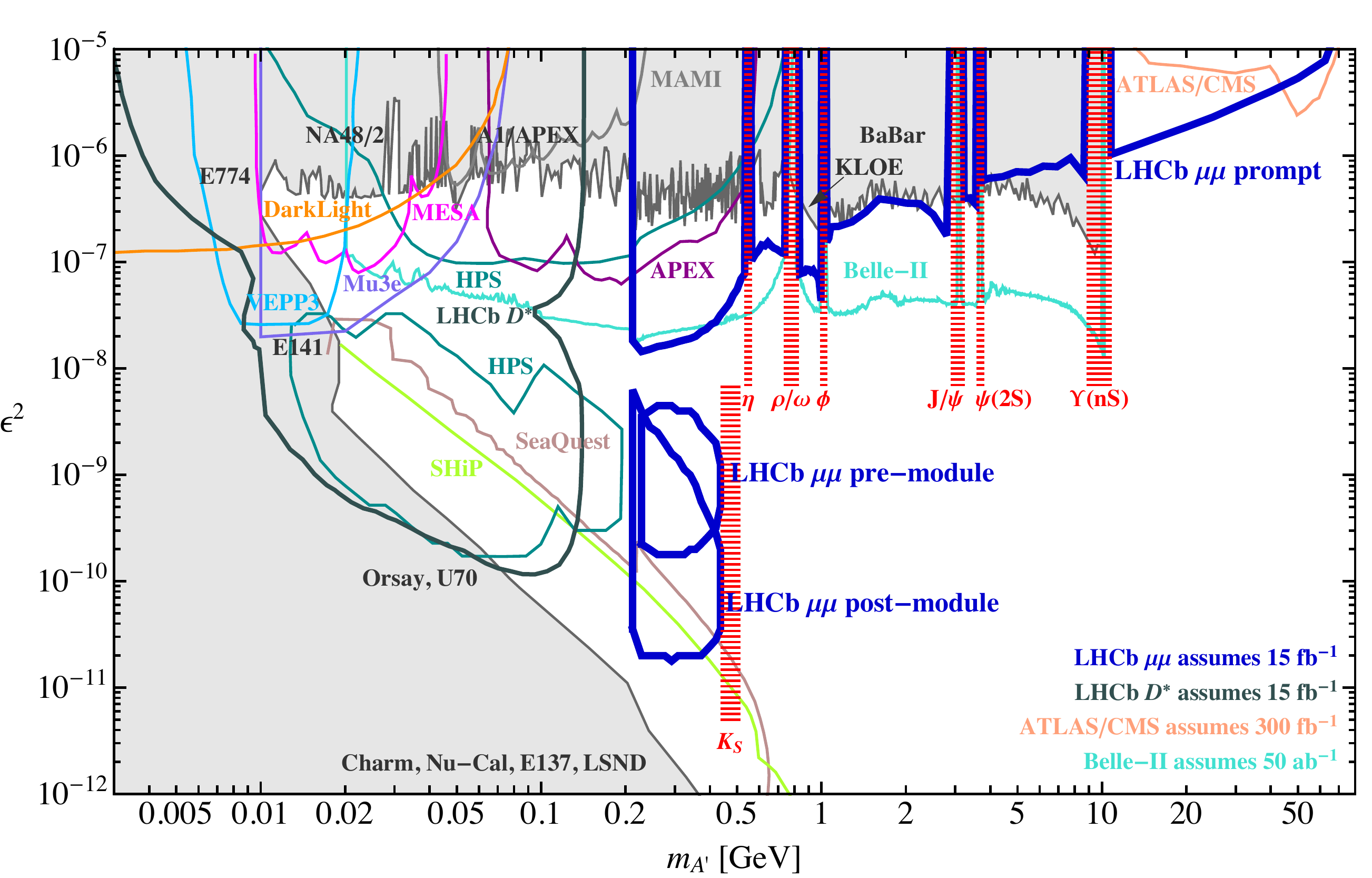}
\caption{Previous and planned experimental bounds on dark photons (adapted from \Ref{Essig:2013lka}) compared to the anticipated LHCb reach for inclusive $A'$ production in the di-muon channel (see the text for definitions of prompt, pre-module, and post-module).  The red vertical bands indicate QCD resonances which would have to be masked in a complete analysis.  The LHCb $D^*$ anticipated limit comes from \Ref{Ilten:2015hya}, and Belle-II comes from \Ref{Hearty}.
}
\label{fig:finalreachmumu}
\end{figure*}

A particularly compelling dark-force scenario is that of a dark photon $A'$ which has small SM couplings via kinetic mixing with the ordinary photon through the operator $\frac{\epsilon}{2} F'_{\mu \nu} F^{\mu \nu}$~\cite{Okun:1982xi,Galison:1983pa,Holdom:1985ag,Pospelov:2007mp,ArkaniHamed:2008qn,Bjorken:2009mm}.
Previous beam dump~\cite{Bergsma:1985is,Konaka:1986cb,Riordan:1987aw,Bjorken:1988as,Bross:1989mp,Davier:1989wz,Athanassopoulos:1997er,Astier:2001ck,Adler:2004hp,Bjorken:2009mm,Artamonov:2009sz,Essig:2010gu,Blumlein:2011mv,Gninenko:2012eq,Blumlein:2013cua},
fixed target~\cite{Abrahamyan:2011gv,Merkel:2014avp,Merkel:2011ze},
collider~\cite{Aubert:2009cp,Curtin:2013fra,Lees:2014xha},
and rare meson decay \cite{Bernardi:1985ny,MeijerDrees:1992kd,Archilli:2011zc,Gninenko:2011uv,Babusci:2012cr,Adlarson:2013eza,Agakishiev:2013fwl,Adare:2014mgk,Batley:2015lha,KLOE:2016lwm} experiments have already played a crucial role in constraining the dark photon mass $m_{A'}$ and kinetic-mixing strength~$\epsilon^2$. Large regions of the $m_{A'}$--$\epsilon^2$ plane, however, are still unexplored (see Fig.~\ref{fig:finalreachmumu}). Looking to the future, a wide variety of innovative experiments have been proposed to further probe the dark photon parameter space~\cite{Essig:2010xa,
Freytsis:2009bh,Balewski:2013oza,
Wojtsekhowski:2012zq,
Beranek:2013yqa,
Echenard:2014lma,
Battaglieri:2014hga,
Curtin:2014cca,
Alekhin:2015byh,
Gardner:2015wea,
Ilten:2015hya},
though new ideas are needed to test $m_{A'} > 2 m_\mu$ and $\epsilon^2 \in [10^{-7}, 10^{-11}]$.

In this Letter, we propose a search for dark photons via the decay
\be
	A' \to\mu^+ \mu^-  \, ,
\ee
at the LHCb experiment during LHC Run~3 (scheduled for 2021--2023).  The potential of LHCb to discover dark photons was recently emphasized in \Ref{Ilten:2015hya}, which exploits the exclusive charm decay mode $D^*\to D^0 A'$ with $A' \to e^+ e^-$.  Here, we consider an inclusive approach where the production mode of $A'$ need not be specified.  An important feature of this search is that it can be made fully data driven, since the $A'$ signal rate can be inferred from measurements of the SM prompt $\mu^+ \mu^-$ spectrum. 
The excellent invariant-mass and vertex resolution of the LHCb detector, along with its unique particle-identification and real-time data-analysis capabilities~\cite{Benson:2015yzo,Aaij:2015bpa}, make it highly sensitive to $A' \to \mu^+ \mu^-$.  We derive the LHCb sensitivity for both prompt  and displaced $A'$ decays, and show that LHCb can probe otherwise inaccessible regions of the $m_{A'}$--$\epsilon^2$ plane.

The $A'$ is a hypothetical massive spin-1 particle that, after electroweak symmetry breaking and diagonalizing the gauge kinetic terms, has a suppressed coupling to the electromagnetic~(EM) current $J^\mu_{\rm EM}$ \cite{Okun:1982xi,Galison:1983pa,Holdom:1985ag,Pospelov:2007mp,ArkaniHamed:2008qn,Bjorken:2009mm}:
\begin{align} 
	\label{eq:LgammaAp}
	\cL_{\gamma A^\prime}  \supset&	
	- \frac{1}{4}F'_{\mu\nu}F^{\prime\mu\nu} 
	+ \frac{1}{2} m^2_{A^\prime} A^{\prime \mu} A^{\prime}_{\mu} 
	+ \epsilon \, e \,  A^\prime_\mu J^\mu_{\rm EM} \, .
\end{align}
There is also a model-dependent coupling to the weak $Z$ current (see e.g.\ \Ref{Barello:2015bhq}), which appears at $\cO(m^2_{A'}/m^2_Z)$.  
We provide nearly model-independent sensitivity estimates for the mass range $m_{A'}\lesssim 10\gev$ by ignoring the coupling to the $Z$.  We include model-dependent $Z$-mixing effects for $m_{A'}\gtrsim 10\gev$, adopting the parameters of \Refs{Cassel:2009pu,Cline:2014dwa}.  


The partial widths of $A'$ to SM leptons are 
\be
	\label{eq:Ap2ll}
	\Gamma_{A'\to \ell^+\ell^-}
=	\tfrac{\epsilon^2 \alpha_{\rm EM}}{3} m_{A'} \left(  1 +2\tfrac{m^2_\ell}{m^2_{A'}} \right) \sqrt{ 1 - 4\tfrac{m^2_\ell}{m^2_{A'}} } \, ,
\ee
where $\ell=e,\mu,\tau$ and $m_{A'} > 2m_{\ell}$.  
Because the $A'$ couples to $J^\mu_{\rm EM}$, the branching fraction of $A'$ to SM hadrons can be extracted from the measured value of $\cR_\mu \equiv \sigma_{e^+e^-\to {\rm hadrons}} / \sigma_{e^+e^-\to \mu^+\mu^-}$ (taken from \Ref{Agashe:2014kda})
\be
	\label{eq:Ap2had}
	\Gamma_{A' \to {\rm hadrons} }
= 	\Gamma_{A'\to \mu^+\mu^-} \cR_{\mu} (m^2_{A'}) \, .
\ee
In particular, \eqref{eq:Ap2had} already includes the effect of the $A'$ mixing with the QCD vector mesons $\rho$, $\omega$, $\phi$, etc.  It is also possible for the $A'$ to couple to non-SM particles with an invisible decay width $\Gamma_{A' \to {\rm invisible}}$, in which case the total $A'$ width is
\be
\Gamma_{A'}=\sum_\ell \Gamma_{A'\to \ell^+\ell^-} +\Gamma_{A' \to {\rm hadrons} } +\Gamma_{A' \to {\rm invisible} } \, .
\ee
Below, we consider $\Gamma_{A' \to {\rm invisible}}=0$, though our analysis can be easily adapted to handle non-vanishing invisible decay modes.

To estimate the $A' \to \mu^+\mu^-$ signal rate, we follow the strategy outlined in \Ref{Bjorken:2009mm}. Consider the signal production process in proton-proton ($pp$) collisions
\be
	S: \quad  pp \to X A' \to X \mu^+\mu^-,
\ee
where $X$ is any (multiparticle) final state. Ignoring $\cO(m^2_{A'}/m^2_Z)$ and $\cO(\alpha_{\rm EM})$ corrections, this process has the identical cross section to the prompt SM process which originates from the EM current
\be
	B_{\rm EM}: \quad  pp \to X \gamma^* \to X \mu^+\mu^-,
\ee
up to differences between the $A'$ and $\gamma^*$ propagators and the kinetic-mixing suppression. Interference between $S$ and $B_{\rm EM}$ is negligible for a narrow $A'$ resonance. Therefore, for \emph{any} selection criteria on $X$, $\mu^+$, and $\mu^-$, the ratio between the differential cross sections is
\be
	\label{eq:diffprediction}
	\frac{\text{d} \sigma_{pp \to X A'\to X \mu^+\mu^-}}{\text{d} \sigma_{pp \to X \gamma^*\to X\mu^+\mu^-}}
=	\epsilon^4\frac{m^4_{\mu\mu}}{(m^2_{\mu\mu} - m^2_{A'} )^2 + \Gamma^2_{A'}m^2_{A'}} \, ,
\ee
where $m_{\mu\mu}$ is the di-muon invariant mass, for the case $\Gamma_{A'} \ll |m_{\mu\mu}-m_{A'}| \ll m_{A'}$.  The $\epsilon^4$ factor arises because both the $A'$ production and decay rates scale like $\epsilon^2$.   

To obtain a signal event count, we integrate over an invariant-mass range of $|m_{\mu \mu} - m_{A'}| < 2\sigma_{m_{\mu\mu}}$, where $\sigma_{m_{\mu\mu}}$ is the detector resolution on $m_{\mu\mu}$.  
The ratio of signal events to prompt EM background events is 
\be
\label{eq:SBgen}
	\frac{S}{B_{\rm EM}} \approx \epsilon^4 \frac{\pi}{8} \frac{ m^2_{A'}}{\Gamma_{A'}\sigma_{m_{\mu\mu}}} \approx \frac{3\pi}{8} \frac{m_{A'}}{\sigma_{m_{\mu\mu}}}  \frac{\epsilon^2 }{\alpha_{\rm EM}(N_\ell+\cR_\mu)},
\ee
neglecting phase space factors for $N_\ell$ leptons lighter than $m_{A'}/2$.  This expression already accounts for the $A' \to \mu^+\mu^-$
branching-fraction suppression when $\cR_\mu$ is large.  Despite the factor of $\epsilon^4$ in \eqref{eq:diffprediction}, the ratio in \eqref{eq:SBgen} is proportional to $\epsilon^2$ because of the $\epsilon^2$ scaling of $\Gamma_{A'}$.

We emphasize that~\eqref{eq:SBgen} holds for \emph{any} final state $X$ (and any kinematic selection) in the $m_{A'}\ll m_Z$ limit for tree-level single-photon processes. In particular, it already includes $\mu^+ \mu^-$ production from QCD vector mesons that mix with the photon.  This allows us to perform a fully data-driven analysis, since the efficiency and acceptance for the (measured) prompt SM process is the same as for the (inferred) signal process, excluding $A'$ lifetime-based effects.  The dominant component of $B_{\rm EM}$ at small $m_{A'}$ comes from meson decays $M\to \mu^+\mu^-Y$, 
especially $\eta \to \mu^+\mu^-\gamma$, and is denoted as $B_M$ (which includes feed-down contributions from heavier meson decays).   There are also two other important components: final state radiation~(FSR) and Drell-Yan~(DY) production.  Non-prompt $\gamma^*$ production is small and only considered as a background.

Beyond $B_{\rm EM}$, there are other important sources of backgrounds that contribute to the reconstructed prompt di-muon sample, ordered by their relative size:
\begin{itemize}
\item $B^{\pi\pi}_{\rm misID}$:  Two pions (and more rarely a kaon and pion) can be misidentified (misID) as a fake di-muon pair, including the contribution from in-flight decays.  This background can be deduced and subtracted in a data-driven way using prompt same-sign di-muon candidates~\cite{ALICE:2011ad,SUP}.
\item $B^{\pi\mu}_{\rm misID}$:  A fake di-muon pair can also arise from one real muon (primarily from charm or beauty decays) combined with one misID pion or kaon.  This background can be subtracted similarly to $B^{\pi\pi}_{\rm misID}$.
\item $B_{\rm BH}$:  The Bethe-Heitler (BH) background played an important role in the analysis of \Ref{Bjorken:2009mm}.  This is a subdominant process at the LHC due in part to the small effective photon luminosity function.  We verified that $B_{\rm BH}$ is small using a parton shower generator (see below), and it will be neglected in estimating the reach.
\end{itemize}
True displaced di-muon pairs, which arise from beauty decays, are rarely reconstructed as prompt at LHCb.  Such backgrounds, however, are dominant in the displaced search discussed below.

Summarizing, the reconstructed prompt di-muon sample contains the following background components:
\be
B_{\rm prompt} = \underbrace{B_M + B_{\rm FSR} + B_{\rm DY}}_{B_{\rm EM}} + \underbrace{B^{\pi\pi}_{\rm misID} + B^{\pi\mu}_{\rm misID}}_{B_{\rm misID}},
\ee
where for simplicity we ignore interference terms between the various $B_{\rm EM}$ components. 
After subtracting $B_{\rm misID}$ from $B_{\rm prompt}$~\cite{ALICE:2011ad,SUP}, we can use \eqref{eq:SBgen} to infer $S$ from $B_{\rm EM}$ for any $m_{A'}$ and $\epsilon^2$.   Since both $B_{\rm prompt}$ and $B_{\rm misID}$ are extracted from data, this strategy is fully data driven.  

We now present an inclusive search strategy for dark photons at LHCb.  The LHCb experiment will upgrade to a triggerless detector-readout system for Run~3 of the LHC~\cite{LHCb-TDR-016}, making it highly efficient at selecting $A' \to \mu^+ \mu^-$ decays in real time.
Therefore, we focus on Run~3 and assume
an integrated luminosity of (see~\Ref{Ilten:2015hya})
\be
	\label{eq:Lum}
	\int \mathcal{L} \, \mathrm{d} t = 15~\text{fb}^{-1}.
\ee
The trigger system currently employed by LHCb is efficient for many $A' \to \mu^+\mu^-$ decays included in our search.
We estimate that the sensitivity in Run~2 will be equivalent to using about 10\% of the data collected in Run~3.
Therefore, inclusion of Run~2 data will not greatly impact the reach by the end of Run~3, though a Run~2 analysis could explore much of the same $m_{A'}-\epsilon^2$ parameter space in the next few years.

The LHCb detector is a forward spectrometer covering the pseudorapidity range $2 < \eta < 5$~\cite{Alves:2008zz,Aaij:2014jba}. 
Within this acceptance, muons with three-momentum $p > 5\gev$ are reconstructed with near 100\% efficiency with a momentum resolution of $\sigma_p / p \approx 0.5\%$ and a di-muon invariant mass resolution of~\cite{Aaij:2014jba,Aaij:2012rt} 
\begin{align}
	\sigma_{m_{\mu\mu}} \approx& \left\{ \begin{matrix}
	4\mev & m_{\mu\mu}<1\gev \\
	0.4\%\, m_{\mu\mu} &m_{\mu\mu}>1\gev
	\end{matrix}\right. \, .
\end{align}
For the displaced $A'$ search, the vertex resolution of LHCb depends on the Lorentz boost factor of the $A'$; we therefore use an event-by-event selection criteria in the analysis below.  That said, it is a reasonable approximation to use a fixed $A'$ proper-lifetime resolution~\cite{Aaij:2014jba}
\be
	\label{eq:sigmatau}
	\sigma_\tau \approx 50~\text{fs}\,,
\ee
except near the di-muon threshold where the opening angle between the muons is small.

To suppress fake muons, our strategy requires muon candidates have (transverse) momenta ($p_T > 0.5\gev$) $p > 10\gev$, and are selected by a neural-network muon-identification algorithm~\cite{Aaij:2014azz} with a muon efficiency of $\epsilon^2_\mu\approx0.50$ and a pion fake rate of $\epsilon^2_{\pi}\approx10^{-6}$~\cite{SUP}.  To a good approximation, the neural-network performance is independent of the kinematics.  Such a low pion misID rate is a unique feature of LHCb and is vital for probing the low-$m_{A'}$ region in $A' \to \mu^+\mu^-$ decays.

To further suppress $B_{\rm misID}$ for $m_{A'}>m_{\phi} \simeq 1.0\gev$, we require muons to satisfy an isolation criterion based on clustering the charged component of the final state with the anti-$k_T$ jet algorithm \cite{Cacciari:2008gp} with $R = 0.5$ in \textsc{FastJet} 3.1.2~\cite{Cacciari:2011ma}; muons with $p_T(\mu)/p_T({\rm jet}) < 0.85$ are rejected, excluding the contribution to $p_T({\rm jet})$ from the other muon if it is contained in the same jet.  By considering charged particles only, this isolation strategy is robust to pileup.  The di-muon isolation efficiencies obtained from simulated LHCb data (see below) are 50\% for FSR, DY, and BH, 25\% for meson decays (dominantly from charmonium states), and 1\% for fake pions ($\pi\pi$ and $\pi\mu$ have similar efficiencies). 

The baseline selection for the LHCb inclusive $A'$ search is therefore:
\begin{enumerate}
\item two opposite-sign muons with $\eta(\mu^\pm) \in [2,5]$, $p(\mu^\pm) > 10~\GeV$, and $p_T(\mu^\pm) > 0.5~\GeV$;
\item a reconstructed $A' \to \mu^+ \mu^-$ candidate with $\eta(A') \in [2,5]$, $p_T(A') > 1~\GeV$, and passing the isolation criterion for $m_{A'}>m_{\phi}$;
\item an $A' \to \mu^+\mu^-$ decay topology consistent with either a prompt or displaced $A'$ decay~\cite{Ilten:2015hya,SUP}.
\end{enumerate}
Following a similar strategy to \Ref{Ilten:2015hya}, we use the reconstructed muon impact parameter~(IP) and $A'$ transverse flight distance $\ell_{\rm T}$ to define three non-overlapping search regions:
\begin{enumerate}
\item {\bf Prompt}: ${\rm IP}_{\mu^\pm} < 2.5\,\sigma_{{\rm IP}}$;
\item {\bf Displaced (pre-module)}:  $\ell_{\rm T} \in [5\,\sigma_{\ell_{\rm T}}, 6\,\text{mm}]$;
\item {\bf Displaced (post-module)}: $\ell_{\rm T} \in [6\,\text{mm}, 22\,\text{mm}]$.
\end{enumerate}
The resolution on IP and $\ell_{\rm T}$ are taken from~\Refs{LHCb-TDR-013,SUP}.
The displaced $A'$ search is restricted to $\ell_{\rm T} < 22\,\text{mm}$ to ensure at least three hits per track in the vertex locator (VELO). 
We define two search regions based on the average $\ell_{\rm T}$ to the first VELO module (i.e.~6\,\text{mm}), where each VELO module is a planar silicon-pixel detector oriented perpendicular to the LHC beamline.   

To estimate the reach for this $A'$ search using the data-driven strategy in \eqref{eq:SBgen}, we need to know $B_{\rm prompt}(m_{\mu\mu})$ with the above selection criteria applied.  To our knowledge, LHCb has not published such a spectrum, so we use \textsc{Pythia}~8.212~\cite{Sjostrand:2014zea} for illustrative purposes to understand the various components of $B_{\rm EM}$.\footnote{We caution the reader that the di-muon spectra published by ATLAS~\cite{Aad:2014zya} and CMS~\cite{Chatrchyan:2012xi} do not impose prompt selection criteria nor do they subtract fake di-muons.  
}  
LHCb has published measurements of $\phi$ meson~\cite{Aaij:2011uk}, charmonium~\cite{Aaij:2011jh}, bottomonium~\cite{LHCb:2012aa}, and DY~\cite{LHCb:2012fja} production in 7\tev $pp$ collisions, and we find that \textsc{Pythia} accurately reproduces these measurements. Therefore, we assume that \textsc{Pythia} also adequately predicts their production  at 14\tev.
The ALICE collaboration has published the low-mass di-muon spectrum at $\sqrt{s} = 7\tev$ in a similar kinematic region as proposed for this search~\cite{ALICE:2011ad}.  Within the kinematic region used by ALICE, we find that \textsc{Pythia} accurately describes the production of the $\eta^{(\prime)}$ mesons, but overestimates $\omega$ and $\rho$ production by factor of two; we therefore reduce the \textsc{Pythia} prediction for these mesons to match the observed ALICE spectrum.   


\begin{figure}
\includegraphics[width=\columnwidth]{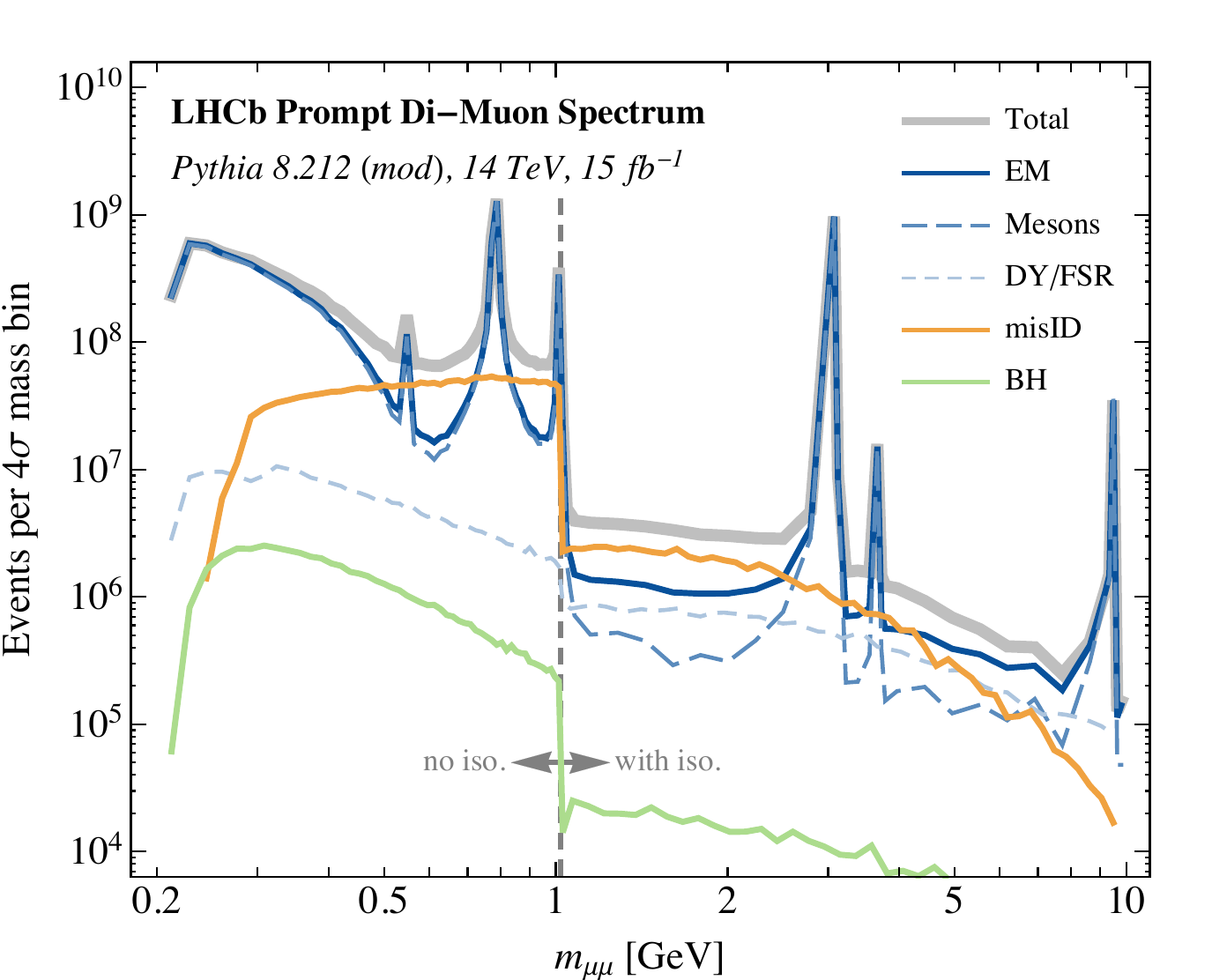}
\caption{Predicted reconstructed di-muon invariant mass spectrum with our prompt selection criteria applied after Run~3, including the isolation criteria for $m_{\mu\mu} > m_\phi$.  ``EM'' denotes the sum of ``Mesons'' and ``DY/FSR''. 
}  
\label{fig:dimuonspectrum}
\end{figure}

Including our selection criteria and modifications, the prompt di-muon spectrum from \textsc{Pythia} is shown in \Fig{fig:dimuonspectrum}. 
The $B_{\rm EM}$ background is dominated by meson decays like $\eta \to \mu^+ \mu^- \gamma$ at low invariant mass, and transitions to DY production $pp \to \gamma^* \to \mu^+ \mu^-$ at larger $m_{\mu\mu}$, with FSR being subdominant throughout.  Note the sharp change in the spectrum at $m_{\mu \mu} = m_{\phi}$ due to the muon-isolation requirement.  We also show in  \Fig{fig:dimuonspectrum} the expected non-EM background contamination from $B_{\rm misID}$ and $B_{\rm BH}$. The misidentification background is large and dominates for $m_{A'} \in [1,3] \gev$, though this is also the region where \textsc{Pythia} likely underestimates di-muon production from excited meson decays (e.g.~$\rho(1450) \to \mu^+ \mu^-$) \cite{SUP}.

We also use \textsc{Pythia} to understand the backgrounds for the displaced $A'$ searches, where the dominant contribution comes from double semi-leptonic heavy-flavor decays of the form $b \to c \, \mu^{\pm}X$ followed by $c \to \mu^{\mp}Y$.  Such decays are highly suppressed by our consistent-decay-topology requirements~\cite{SUP}, but they still contribute at a large rate because of the copious heavy-flavor production in high-energy $pp$ collisions. Semi-leptonic decays of charm and beauty mesons, where one real muon and one fake muon arise from the same secondary vertex, also contribute but at a much lower rate.  Decays of heavy-flavor hadrons with two misID pions or with $\gamma^* \to \mu^+\mu^-$ are similarly subdominant.

For the pre-module displaced region, we find $\approx 10^4$ background events per $\pm2\sigma_{m_{\mu\mu}}$ mass bin.  
For the post-module displaced region, relevant for long-lived dark photons with $\tau_{A'}\gg\tau_{D,B}$, 
we estimate the background to be $\approx 25$ candidates per mass bin by scaling the observed combinatorial background in a published LHCb $K_S \to \mu^+\mu^-$ search~\cite{Aaij:2014azz} by the increase in luminosity used in this analysis.
In the post-module region, the heavy-flavor background is on the order of few events per bin, and the dominant contribution is from interactions with the detector material.  This contribution can likely be reduced following a strategy similar to \Ref{Ilten:2015hya}.

The estimated sensitivity of LHCb to inclusive $A'$ production is shown in \Fig{fig:finalreachmumu}.  For the prompt $A'$ search, we estimate $S$ from $B_{\rm EM}$ using data in the neighboring sidebands and take $S/\sqrt{B_{\rm prompt}} \approx 2$ as a rough criterion for the exclusion limit.  This sideband method fails near narrow QCD resonances, which would need a dedicated analysis. 
 Figure~\ref{fig:finalreachmumu} shows that for $m_{A'} \in [2m_\mu,m_\phi]$ one can probe $\epsilon^2$ down to $10^{-8}$--$10^{-7}$ with the prompt search, improving on current limits.  The reach is limited at higher masses due to $B_{\rm misID}$, where the expected sensitivity is comparable to the present bound.  Going to higher masses where the $A'$ production rate depends on model-dependent mixing with the $Z$, LHCb can extend anticipated ATLAS and CMS limits \cite{Curtin:2014cca} for $m_{A'} \in [10,40]\gev$.
 
For the displaced $A'$ search, the spectrum of $A'$ Lorentz boost factors $\gamma_{\mu \mu} \equiv E_{\mu \mu} / m_{\mu \mu}$ can be inferred from the prompt $\gamma^* \to \ell^+ \ell^-$ spectrum observed in data in a given $m_{\mu\mu}$ bin; the $A'$ lifetime acceptance can then be obtained from simulation.   Following the background discussion above, the exclusion criterion for the pre-module (post-module) search is $S\approx 2 \sqrt{B} \approx 200$ ($S\approx 2 \sqrt{B} \approx 10$), yielding the regions shown in \Fig{fig:finalreachmumu}.   A comparable reach is obtained by simply assuming the fixed proper-lifetime resolution in \eqref{eq:sigmatau}.  Because of the large $\eta \to \gamma A'$ rate, the displaced search has the potential to probe $m_{A'} \in [2m_\mu,m_\eta]$ with $\epsilon\in[10^{-11},10^{-8}]$, a region that is challenging to access through other experiments.  

There are a number of possible improvements and generalizations to this $A'$ search.  For example, dark photons can be searched for during LHC Run~2, by adapting our analysis to include di-muon hardware trigger requirements.  Because the search is entirely data driven, di-muon triggers need not be fully efficient to be useful in such an analysis.  The real-time analysis, event-selection, and multi-search-region~\cite{Williams:2015xfa} strategies employed by LHCb could be improved, and data collected in LHC Runs 4 and 5 would greatly improve the sensitivity~\cite{SUP}.  One could also pursue a semi-inclusive strategy, where an $A'$ candidate is selected along with another required object; for most semi-inclusive modes, one can still use the data-driven method in \eqref{eq:SBgen}.  If the fake muon backgrounds could be controlled, a similar search could be performed at ATLAS and CMS.   Beyond dark photons, these searches are sensitive to spin-0 di-muon resonances~(see related work in \cite{Freytsis:2009ct,Aaij:2015tna,Haisch:2016hzu}).  
An inclusive $A'$ search in the electron channel could explore the $m_{A'} \in [2m_e,2m_\mu]$ mass region, though this is considerably more challenging due to  Bremsstrahlung radiation and multiple scattering \cite{SUP}.

In summary, we proposed an inclusive search strategy for dark photons at the LHCb experiment using di-muon resonances.  Since the coupling of the $A'$ to the standard model is dictated by the kinetic-mixing parameter $\epsilon^2$, the signal rate can be directly inferred from the off-shell photon rate, enabling a data-driven search. Through a combination of prompt and displaced searches, LHCb is sensitive to interesting regions in the $m_{A'}$--$\epsilon^2$ parameter space, some of which are difficult to probe with other proposed experiments.  This search leverages the excellent invariant-mass and vertex resolution of LHCb, along with its unique particle-identification and real-time data-analysis capabilities.  Provided that the appropriate real-time selections are employed starting this year, LHCb could probe much of this parameter space using data collected in Run~2 of the LHC.  Given the simplicity of this proposed search strategy, it could easily be adapted to other experiments at the LHC and beyond.

\begin{acknowledgments}
We thank R.~Essig, T.~Gershon, J.~Kamenik, Z.~Ligeti, and V.~Vagnoni for helpful feedback. 
Y.S., J.T., and W.X. are supported by the U.S. Department of Energy (DOE) under cooperative research agreement DE-SC-00012567.  J.T. is also supported by the DOE Early Career research program DE-SC-0006389, and by a Sloan Research Fellowship from the Alfred P. Sloan Foundation.  M.W. and P.I. are supported by the U.S.\ National Science Foundation grant PHY-1306550.
\end{acknowledgments}

\bibliography{ref}

\begin{thebibliography}{79}%
\makeatletter
\providecommand \@ifxundefined [1]{%
 \@ifx{#1\undefined}
}%
\providecommand \@ifnum [1]{%
 \ifnum #1\expandafter \@firstoftwo
 \else \expandafter \@secondoftwo
 \fi
}%
\providecommand \@ifx [1]{%
 \ifx #1\expandafter \@firstoftwo
 \else \expandafter \@secondoftwo
 \fi
}%
\providecommand \natexlab [1]{#1}%
\providecommand \enquote  [1]{``#1''}%
\providecommand \bibnamefont  [1]{#1}%
\providecommand \bibfnamefont [1]{#1}%
\providecommand \citenamefont [1]{#1}%
\providecommand \href@noop [0]{\@secondoftwo}%
\providecommand \href [0]{\begingroup \@sanitize@url \@href}%
\providecommand \@href[1]{\@@startlink{#1}\@@href}%
\providecommand \@@href[1]{\endgroup#1\@@endlink}%
\providecommand \@sanitize@url [0]{\catcode `\\12\catcode `\$12\catcode
  `\&12\catcode `\#12\catcode `\^12\catcode `\_12\catcode `\%12\relax}%
\providecommand \@@startlink[1]{}%
\providecommand \@@endlink[0]{}%
\providecommand \url  [0]{\begingroup\@sanitize@url \@url }%
\providecommand \@url [1]{\endgroup\@href {#1}{\urlprefix }}%
\providecommand \urlprefix  [0]{URL }%
\providecommand \Eprint [0]{\href }%
\providecommand \doibase [0]{http://dx.doi.org/}%
\providecommand \selectlanguage [0]{\@gobble}%
\providecommand \bibinfo  [0]{\@secondoftwo}%
\providecommand \bibfield  [0]{\@secondoftwo}%
\providecommand \translation [1]{[#1]}%
\providecommand \BibitemOpen [0]{}%
\providecommand \bibitemStop [0]{}%
\providecommand \bibitemNoStop [0]{.\EOS\space}%
\providecommand \EOS [0]{\spacefactor3000\relax}%
\providecommand \BibitemShut  [1]{\csname bibitem#1\endcsname}%
\let\auto@bib@innerbib\@empty
\bibitem [{\citenamefont {Essig}\ \emph {et~al.}(2013)\citenamefont {Essig}
  \emph {et~al.}}]{Essig:2013lka}%
  \BibitemOpen
  \bibfield  {author} {\bibinfo {author} {\bibfnamefont {R.}~\bibnamefont
  {Essig}} \emph {et~al.},\ }in\ \href
  {http://inspirehep.net/record/1263039/files/arXiv:1311.0029.pdf} {\emph
  {\bibinfo {booktitle} {{Community Summer Study 2013: Snowmass on the
  Mississippi (CSS2013) Minneapolis, MN, USA, July 29-August 6, 2013}}}}\
  (\bibinfo {year} {2013})\ \Eprint {http://arxiv.org/abs/1311.0029}
  {arXiv:1311.0029 [hep-ph]} \BibitemShut {NoStop}%
\bibitem [{\citenamefont {Okun}(1982)}]{Okun:1982xi}%
  \BibitemOpen
  \bibfield  {author} {\bibinfo {author} {\bibfnamefont {L.~B.}\ \bibnamefont
  {Okun}},\ }\href@noop {} {\bibfield  {journal} {\bibinfo  {journal} {Sov.
  Phys. JETP}\ }\textbf {\bibinfo {volume} {56}},\ \bibinfo {pages} {502}
  (\bibinfo {year} {1982})},\ \bibinfo {note} {[Zh. Eksp. Teor.
  Fiz.83,892(1982)]}\BibitemShut {NoStop}%
\bibitem [{\citenamefont {Galison}\ and\ \citenamefont
  {Manohar}(1984)}]{Galison:1983pa}%
  \BibitemOpen
  \bibfield  {author} {\bibinfo {author} {\bibfnamefont {P.}~\bibnamefont
  {Galison}}\ and\ \bibinfo {author} {\bibfnamefont {A.}~\bibnamefont
  {Manohar}},\ }\href {\doibase 10.1016/0370-2693(84)91161-4} {\bibfield
  {journal} {\bibinfo  {journal} {Phys. Lett.}\ }\textbf {\bibinfo {volume}
  {B136}},\ \bibinfo {pages} {279} (\bibinfo {year} {1984})}\BibitemShut
  {NoStop}%
\bibitem [{\citenamefont {Holdom}(1986)}]{Holdom:1985ag}%
  \BibitemOpen
  \bibfield  {author} {\bibinfo {author} {\bibfnamefont {B.}~\bibnamefont
  {Holdom}},\ }\href {\doibase 10.1016/0370-2693(86)91377-8} {\bibfield
  {journal} {\bibinfo  {journal} {Phys. Lett.}\ }\textbf {\bibinfo {volume}
  {B166}},\ \bibinfo {pages} {196} (\bibinfo {year} {1986})}\BibitemShut
  {NoStop}%
\bibitem [{\citenamefont {Pospelov}\ \emph {et~al.}(2008)\citenamefont
  {Pospelov}, \citenamefont {Ritz},\ and\ \citenamefont
  {Voloshin}}]{Pospelov:2007mp}%
  \BibitemOpen
  \bibfield  {author} {\bibinfo {author} {\bibfnamefont {M.}~\bibnamefont
  {Pospelov}}, \bibinfo {author} {\bibfnamefont {A.}~\bibnamefont {Ritz}}, \
  and\ \bibinfo {author} {\bibfnamefont {M.~B.}\ \bibnamefont {Voloshin}},\
  }\href {\doibase 10.1016/j.physletb.2008.02.052} {\bibfield  {journal}
  {\bibinfo  {journal} {Phys. Lett.}\ }\textbf {\bibinfo {volume} {B662}},\
  \bibinfo {pages} {53} (\bibinfo {year} {2008})},\ \Eprint
  {http://arxiv.org/abs/0711.4866} {arXiv:0711.4866 [hep-ph]} \BibitemShut
  {NoStop}%
\bibitem [{\citenamefont {Arkani-Hamed}\ \emph {et~al.}(2009)\citenamefont
  {Arkani-Hamed}, \citenamefont {Finkbeiner}, \citenamefont {Slatyer},\ and\
  \citenamefont {Weiner}}]{ArkaniHamed:2008qn}%
  \BibitemOpen
  \bibfield  {author} {\bibinfo {author} {\bibfnamefont {N.}~\bibnamefont
  {Arkani-Hamed}}, \bibinfo {author} {\bibfnamefont {D.~P.}\ \bibnamefont
  {Finkbeiner}}, \bibinfo {author} {\bibfnamefont {T.~R.}\ \bibnamefont
  {Slatyer}}, \ and\ \bibinfo {author} {\bibfnamefont {N.}~\bibnamefont
  {Weiner}},\ }\href {\doibase 10.1103/PhysRevD.79.015014} {\bibfield
  {journal} {\bibinfo  {journal} {Phys. Rev.}\ }\textbf {\bibinfo {volume}
  {D79}},\ \bibinfo {pages} {015014} (\bibinfo {year} {2009})},\ \Eprint
  {http://arxiv.org/abs/0810.0713} {arXiv:0810.0713 [hep-ph]} \BibitemShut
  {NoStop}%
\bibitem [{\citenamefont {Bjorken}\ \emph {et~al.}(2009)\citenamefont
  {Bjorken}, \citenamefont {Essig}, \citenamefont {Schuster},\ and\
  \citenamefont {Toro}}]{Bjorken:2009mm}%
  \BibitemOpen
  \bibfield  {author} {\bibinfo {author} {\bibfnamefont {J.~D.}\ \bibnamefont
  {Bjorken}}, \bibinfo {author} {\bibfnamefont {R.}~\bibnamefont {Essig}},
  \bibinfo {author} {\bibfnamefont {P.}~\bibnamefont {Schuster}}, \ and\
  \bibinfo {author} {\bibfnamefont {N.}~\bibnamefont {Toro}},\ }\href {\doibase
  10.1103/PhysRevD.80.075018} {\bibfield  {journal} {\bibinfo  {journal} {Phys.
  Rev.}\ }\textbf {\bibinfo {volume} {D80}},\ \bibinfo {pages} {075018}
  (\bibinfo {year} {2009})},\ \Eprint {http://arxiv.org/abs/0906.0580}
  {arXiv:0906.0580 [hep-ph]} \BibitemShut {NoStop}%
\bibitem [{\citenamefont {Bergsma}\ \emph {et~al.}(1986)\citenamefont {Bergsma}
  \emph {et~al.}}]{Bergsma:1985is}%
  \BibitemOpen
  \bibfield  {author} {\bibinfo {author} {\bibfnamefont {F.}~\bibnamefont
  {Bergsma}} \emph {et~al.} (\bibinfo {collaboration} {CHARM}),\ }\href
  {\doibase 10.1016/0370-2693(86)91601-1} {\bibfield  {journal} {\bibinfo
  {journal} {Phys. Lett.}\ }\textbf {\bibinfo {volume} {B166}},\ \bibinfo
  {pages} {473} (\bibinfo {year} {1986})}\BibitemShut {NoStop}%
\bibitem [{\citenamefont {Konaka}\ \emph {et~al.}(1986)\citenamefont {Konaka}
  \emph {et~al.}}]{Konaka:1986cb}%
  \BibitemOpen
  \bibfield  {author} {\bibinfo {author} {\bibfnamefont {A.}~\bibnamefont
  {Konaka}} \emph {et~al.},\ }\bibfield  {booktitle} {\emph {\bibinfo
  {booktitle} {{Proceedings, 23RD International Conference on High Energy
  Physics, JULY 16-23, 1986, Berkeley, CA}}},\ }\href {\doibase
  10.1103/PhysRevLett.57.659} {\bibfield  {journal} {\bibinfo  {journal} {Phys.
  Rev. Lett.}\ }\textbf {\bibinfo {volume} {57}},\ \bibinfo {pages} {659}
  (\bibinfo {year} {1986})}\BibitemShut {NoStop}%
\bibitem [{\citenamefont {Riordan}\ \emph {et~al.}(1987)\citenamefont {Riordan}
  \emph {et~al.}}]{Riordan:1987aw}%
  \BibitemOpen
  \bibfield  {author} {\bibinfo {author} {\bibfnamefont {E.~M.}\ \bibnamefont
  {Riordan}} \emph {et~al.},\ }\href {\doibase 10.1103/PhysRevLett.59.755}
  {\bibfield  {journal} {\bibinfo  {journal} {Phys. Rev. Lett.}\ }\textbf
  {\bibinfo {volume} {59}},\ \bibinfo {pages} {755} (\bibinfo {year}
  {1987})}\BibitemShut {NoStop}%
\bibitem [{\citenamefont {Bjorken}\ \emph {et~al.}(1988)\citenamefont
  {Bjorken}, \citenamefont {Ecklund}, \citenamefont {Nelson}, \citenamefont
  {Abashian}, \citenamefont {Church}, \citenamefont {Lu}, \citenamefont {Mo},
  \citenamefont {Nunamaker},\ and\ \citenamefont {Rassmann}}]{Bjorken:1988as}%
  \BibitemOpen
  \bibfield  {author} {\bibinfo {author} {\bibfnamefont {J.~D.}\ \bibnamefont
  {Bjorken}}, \bibinfo {author} {\bibfnamefont {S.}~\bibnamefont {Ecklund}},
  \bibinfo {author} {\bibfnamefont {W.~R.}\ \bibnamefont {Nelson}}, \bibinfo
  {author} {\bibfnamefont {A.}~\bibnamefont {Abashian}}, \bibinfo {author}
  {\bibfnamefont {C.}~\bibnamefont {Church}}, \bibinfo {author} {\bibfnamefont
  {B.}~\bibnamefont {Lu}}, \bibinfo {author} {\bibfnamefont {L.~W.}\
  \bibnamefont {Mo}}, \bibinfo {author} {\bibfnamefont {T.~A.}\ \bibnamefont
  {Nunamaker}}, \ and\ \bibinfo {author} {\bibfnamefont {P.}~\bibnamefont
  {Rassmann}},\ }\href {\doibase 10.1103/PhysRevD.38.3375} {\bibfield
  {journal} {\bibinfo  {journal} {Phys. Rev.}\ }\textbf {\bibinfo {volume}
  {D38}},\ \bibinfo {pages} {3375} (\bibinfo {year} {1988})}\BibitemShut
  {NoStop}%
\bibitem [{\citenamefont {Bross}\ \emph {et~al.}(1991)\citenamefont {Bross},
  \citenamefont {Crisler}, \citenamefont {Pordes}, \citenamefont {Volk},
  \citenamefont {Errede},\ and\ \citenamefont {Wrbanek}}]{Bross:1989mp}%
  \BibitemOpen
  \bibfield  {author} {\bibinfo {author} {\bibfnamefont {A.}~\bibnamefont
  {Bross}}, \bibinfo {author} {\bibfnamefont {M.}~\bibnamefont {Crisler}},
  \bibinfo {author} {\bibfnamefont {S.~H.}\ \bibnamefont {Pordes}}, \bibinfo
  {author} {\bibfnamefont {J.}~\bibnamefont {Volk}}, \bibinfo {author}
  {\bibfnamefont {S.}~\bibnamefont {Errede}}, \ and\ \bibinfo {author}
  {\bibfnamefont {J.}~\bibnamefont {Wrbanek}},\ }\href {\doibase
  10.1103/PhysRevLett.67.2942} {\bibfield  {journal} {\bibinfo  {journal}
  {Phys. Rev. Lett.}\ }\textbf {\bibinfo {volume} {67}},\ \bibinfo {pages}
  {2942} (\bibinfo {year} {1991})}\BibitemShut {NoStop}%
\bibitem [{\citenamefont {Davier}\ and\ \citenamefont
  {Nguyen~Ngoc}(1989)}]{Davier:1989wz}%
  \BibitemOpen
  \bibfield  {author} {\bibinfo {author} {\bibfnamefont {M.}~\bibnamefont
  {Davier}}\ and\ \bibinfo {author} {\bibfnamefont {H.}~\bibnamefont
  {Nguyen~Ngoc}},\ }\href {\doibase 10.1016/0370-2693(89)90174-3} {\bibfield
  {journal} {\bibinfo  {journal} {Phys. Lett.}\ }\textbf {\bibinfo {volume}
  {B229}},\ \bibinfo {pages} {150} (\bibinfo {year} {1989})}\BibitemShut
  {NoStop}%
\bibitem [{\citenamefont {Athanassopoulos}\ \emph {et~al.}(1998)\citenamefont
  {Athanassopoulos} \emph {et~al.}}]{Athanassopoulos:1997er}%
  \BibitemOpen
  \bibfield  {author} {\bibinfo {author} {\bibfnamefont {C.}~\bibnamefont
  {Athanassopoulos}} \emph {et~al.} (\bibinfo {collaboration} {LSND}),\ }\href
  {\doibase 10.1103/PhysRevC.58.2489} {\bibfield  {journal} {\bibinfo
  {journal} {Phys. Rev.}\ }\textbf {\bibinfo {volume} {C58}},\ \bibinfo {pages}
  {2489} (\bibinfo {year} {1998})},\ \Eprint
  {http://arxiv.org/abs/nucl-ex/9706006} {arXiv:nucl-ex/9706006 [nucl-ex]}
  \BibitemShut {NoStop}%
\bibitem [{\citenamefont {Astier}\ \emph {et~al.}(2001)\citenamefont {Astier}
  \emph {et~al.}}]{Astier:2001ck}%
  \BibitemOpen
  \bibfield  {author} {\bibinfo {author} {\bibfnamefont {P.}~\bibnamefont
  {Astier}} \emph {et~al.} (\bibinfo {collaboration} {NOMAD collaboration}),\
  }\href {\doibase 10.1016/S0370-2693(01)00362-8} {\bibfield  {journal}
  {\bibinfo  {journal} {Phys. Lett.}\ }\textbf {\bibinfo {volume} {B506}},\
  \bibinfo {pages} {27} (\bibinfo {year} {2001})},\ \Eprint
  {http://arxiv.org/abs/hep-ex/0101041} {arXiv:hep-ex/0101041 [hep-ex]}
  \BibitemShut {NoStop}%
\bibitem [{\citenamefont {Adler}\ \emph {et~al.}(2004)\citenamefont {Adler}
  \emph {et~al.}}]{Adler:2004hp}%
  \BibitemOpen
  \bibfield  {author} {\bibinfo {author} {\bibfnamefont {S.}~\bibnamefont
  {Adler}} \emph {et~al.} (\bibinfo {collaboration} {E787}),\ }\href {\doibase
  10.1103/PhysRevD.70.037102} {\bibfield  {journal} {\bibinfo  {journal} {Phys.
  Rev.}\ }\textbf {\bibinfo {volume} {D70}},\ \bibinfo {pages} {037102}
  (\bibinfo {year} {2004})},\ \Eprint {http://arxiv.org/abs/hep-ex/0403034}
  {arXiv:hep-ex/0403034 [hep-ex]} \BibitemShut {NoStop}%
\bibitem [{\citenamefont {Artamonov}\ \emph {et~al.}(2009)\citenamefont
  {Artamonov} \emph {et~al.}}]{Artamonov:2009sz}%
  \BibitemOpen
  \bibfield  {author} {\bibinfo {author} {\bibfnamefont {A.~V.}\ \bibnamefont
  {Artamonov}} \emph {et~al.} (\bibinfo {collaboration} {BNL-E949}),\ }\href
  {\doibase 10.1103/PhysRevD.79.092004} {\bibfield  {journal} {\bibinfo
  {journal} {Phys. Rev.}\ }\textbf {\bibinfo {volume} {D79}},\ \bibinfo {pages}
  {092004} (\bibinfo {year} {2009})},\ \Eprint {http://arxiv.org/abs/0903.0030}
  {arXiv:0903.0030 [hep-ex]} \BibitemShut {NoStop}%
\bibitem [{\citenamefont {Essig}\ \emph {et~al.}(2010)\citenamefont {Essig},
  \citenamefont {Harnik}, \citenamefont {Kaplan},\ and\ \citenamefont
  {Toro}}]{Essig:2010gu}%
  \BibitemOpen
  \bibfield  {author} {\bibinfo {author} {\bibfnamefont {R.}~\bibnamefont
  {Essig}}, \bibinfo {author} {\bibfnamefont {R.}~\bibnamefont {Harnik}},
  \bibinfo {author} {\bibfnamefont {J.}~\bibnamefont {Kaplan}}, \ and\ \bibinfo
  {author} {\bibfnamefont {N.}~\bibnamefont {Toro}},\ }\href {\doibase
  10.1103/PhysRevD.82.113008} {\bibfield  {journal} {\bibinfo  {journal} {Phys.
  Rev.}\ }\textbf {\bibinfo {volume} {D82}},\ \bibinfo {pages} {113008}
  (\bibinfo {year} {2010})},\ \Eprint {http://arxiv.org/abs/1008.0636}
  {arXiv:1008.0636 [hep-ph]} \BibitemShut {NoStop}%
\bibitem [{\citenamefont {Blumlein}\ and\ \citenamefont
  {Brunner}(2011)}]{Blumlein:2011mv}%
  \BibitemOpen
  \bibfield  {author} {\bibinfo {author} {\bibfnamefont {J.}~\bibnamefont
  {Blumlein}}\ and\ \bibinfo {author} {\bibfnamefont {J.}~\bibnamefont
  {Brunner}},\ }\href {\doibase 10.1016/j.physletb.2011.05.046} {\bibfield
  {journal} {\bibinfo  {journal} {Phys. Lett.}\ }\textbf {\bibinfo {volume}
  {B701}},\ \bibinfo {pages} {155} (\bibinfo {year} {2011})},\ \Eprint
  {http://arxiv.org/abs/1104.2747} {arXiv:1104.2747 [hep-ex]} \BibitemShut
  {NoStop}%
\bibitem [{\citenamefont {Gninenko}(2012{\natexlab{a}})}]{Gninenko:2012eq}%
  \BibitemOpen
  \bibfield  {author} {\bibinfo {author} {\bibfnamefont {S.}~\bibnamefont
  {Gninenko}},\ }\href {\doibase 10.1016/j.physletb.2012.06.002} {\bibfield
  {journal} {\bibinfo  {journal} {Phys. Lett.}\ }\textbf {\bibinfo {volume}
  {B713}},\ \bibinfo {pages} {244} (\bibinfo {year} {2012}{\natexlab{a}})},\
  \Eprint {http://arxiv.org/abs/1204.3583} {arXiv:1204.3583 [hep-ph]}
  \BibitemShut {NoStop}%
\bibitem [{\citenamefont {Blümlein}\ and\ \citenamefont
  {Brunner}(2014)}]{Blumlein:2013cua}%
  \BibitemOpen
  \bibfield  {author} {\bibinfo {author} {\bibfnamefont {J.}~\bibnamefont
  {Blümlein}}\ and\ \bibinfo {author} {\bibfnamefont {J.}~\bibnamefont
  {Brunner}},\ }\href {\doibase 10.1016/j.physletb.2014.02.029} {\bibfield
  {journal} {\bibinfo  {journal} {Phys. Lett.}\ }\textbf {\bibinfo {volume}
  {B731}},\ \bibinfo {pages} {320} (\bibinfo {year} {2014})},\ \Eprint
  {http://arxiv.org/abs/1311.3870} {arXiv:1311.3870 [hep-ph]} \BibitemShut
  {NoStop}%
\bibitem [{\citenamefont {Abrahamyan}\ \emph {et~al.}(2011)\citenamefont
  {Abrahamyan} \emph {et~al.}}]{Abrahamyan:2011gv}%
  \BibitemOpen
  \bibfield  {author} {\bibinfo {author} {\bibfnamefont {S.}~\bibnamefont
  {Abrahamyan}} \emph {et~al.} (\bibinfo {collaboration} {APEX
  collaboration}),\ }\href {\doibase 10.1103/PhysRevLett.107.191804} {\bibfield
   {journal} {\bibinfo  {journal} {Phys. Rev. Lett.}\ }\textbf {\bibinfo
  {volume} {107}},\ \bibinfo {pages} {191804} (\bibinfo {year} {2011})},\
  \Eprint {http://arxiv.org/abs/1108.2750} {arXiv:1108.2750 [hep-ex]}
  \BibitemShut {NoStop}%
\bibitem [{\citenamefont {Merkel}\ \emph {et~al.}(2014)\citenamefont {Merkel}
  \emph {et~al.}}]{Merkel:2014avp}%
  \BibitemOpen
  \bibfield  {author} {\bibinfo {author} {\bibfnamefont {H.}~\bibnamefont
  {Merkel}} \emph {et~al.},\ }\href {\doibase 10.1103/PhysRevLett.112.221802}
  {\bibfield  {journal} {\bibinfo  {journal} {Phys. Rev. Lett.}\ }\textbf
  {\bibinfo {volume} {112}},\ \bibinfo {pages} {221802} (\bibinfo {year}
  {2014})},\ \Eprint {http://arxiv.org/abs/1404.5502} {arXiv:1404.5502
  [hep-ex]} \BibitemShut {NoStop}%
\bibitem [{\citenamefont {Merkel}\ \emph {et~al.}(2011)\citenamefont {Merkel}
  \emph {et~al.}}]{Merkel:2011ze}%
  \BibitemOpen
  \bibfield  {author} {\bibinfo {author} {\bibfnamefont {H.}~\bibnamefont
  {Merkel}} \emph {et~al.} (\bibinfo {collaboration} {A1}),\ }\href {\doibase
  10.1103/PhysRevLett.106.251802} {\bibfield  {journal} {\bibinfo  {journal}
  {Phys. Rev. Lett.}\ }\textbf {\bibinfo {volume} {106}},\ \bibinfo {pages}
  {251802} (\bibinfo {year} {2011})},\ \Eprint {http://arxiv.org/abs/1101.4091}
  {arXiv:1101.4091 [nucl-ex]} \BibitemShut {NoStop}%
\bibitem [{\citenamefont {Aubert}\ \emph {et~al.}(2009)\citenamefont {Aubert}
  \emph {et~al.}}]{Aubert:2009cp}%
  \BibitemOpen
  \bibfield  {author} {\bibinfo {author} {\bibfnamefont {B.}~\bibnamefont
  {Aubert}} \emph {et~al.} (\bibinfo {collaboration} {BaBar}),\ }\href
  {\doibase 10.1103/PhysRevLett.103.081803} {\bibfield  {journal} {\bibinfo
  {journal} {Phys. Rev. Lett.}\ }\textbf {\bibinfo {volume} {103}},\ \bibinfo
  {pages} {081803} (\bibinfo {year} {2009})},\ \Eprint
  {http://arxiv.org/abs/0905.4539} {arXiv:0905.4539 [hep-ex]} \BibitemShut
  {NoStop}%
\bibitem [{\citenamefont {Curtin}\ \emph {et~al.}(2014)\citenamefont {Curtin}
  \emph {et~al.}}]{Curtin:2013fra}%
  \BibitemOpen
  \bibfield  {author} {\bibinfo {author} {\bibfnamefont {D.}~\bibnamefont
  {Curtin}} \emph {et~al.},\ }\href {\doibase 10.1103/PhysRevD.90.075004}
  {\bibfield  {journal} {\bibinfo  {journal} {Phys. Rev.}\ }\textbf {\bibinfo
  {volume} {D90}},\ \bibinfo {pages} {075004} (\bibinfo {year} {2014})},\
  \Eprint {http://arxiv.org/abs/1312.4992} {arXiv:1312.4992 [hep-ph]}
  \BibitemShut {NoStop}%
\bibitem [{\citenamefont {Lees}\ \emph {et~al.}(2014)\citenamefont {Lees} \emph
  {et~al.}}]{Lees:2014xha}%
  \BibitemOpen
  \bibfield  {author} {\bibinfo {author} {\bibfnamefont {J.~P.}\ \bibnamefont
  {Lees}} \emph {et~al.} (\bibinfo {collaboration} {BaBar}),\ }\href {\doibase
  10.1103/PhysRevLett.113.201801} {\bibfield  {journal} {\bibinfo  {journal}
  {Phys. Rev. Lett.}\ }\textbf {\bibinfo {volume} {113}},\ \bibinfo {pages}
  {201801} (\bibinfo {year} {2014})},\ \Eprint {http://arxiv.org/abs/1406.2980}
  {arXiv:1406.2980 [hep-ex]} \BibitemShut {NoStop}%
\bibitem [{\citenamefont {Bernardi}\ \emph {et~al.}(1986)\citenamefont
  {Bernardi}, \citenamefont {Carugno}, \citenamefont {Chauveau}, \citenamefont
  {Dicarlo}, \citenamefont {Dris} \emph {et~al.}}]{Bernardi:1985ny}%
  \BibitemOpen
  \bibfield  {author} {\bibinfo {author} {\bibfnamefont {G.}~\bibnamefont
  {Bernardi}}, \bibinfo {author} {\bibfnamefont {G.}~\bibnamefont {Carugno}},
  \bibinfo {author} {\bibfnamefont {J.}~\bibnamefont {Chauveau}}, \bibinfo
  {author} {\bibfnamefont {F.}~\bibnamefont {Dicarlo}}, \bibinfo {author}
  {\bibfnamefont {M.}~\bibnamefont {Dris}},  \emph {et~al.},\ }\href {\doibase
  10.1016/0370-2693(86)91602-3} {\bibfield  {journal} {\bibinfo  {journal}
  {Phys. Lett.}\ }\textbf {\bibinfo {volume} {B166}},\ \bibinfo {pages} {479}
  (\bibinfo {year} {1986})}\BibitemShut {NoStop}%
\bibitem [{\citenamefont {Meijer~Drees}\ \emph {et~al.}(1992)\citenamefont
  {Meijer~Drees} \emph {et~al.}}]{MeijerDrees:1992kd}%
  \BibitemOpen
  \bibfield  {author} {\bibinfo {author} {\bibfnamefont {R.}~\bibnamefont
  {Meijer~Drees}} \emph {et~al.} (\bibinfo {collaboration} {SINDRUM I
  collaboration}),\ }\href {\doibase 10.1103/PhysRevLett.68.3845} {\bibfield
  {journal} {\bibinfo  {journal} {Phys. Rev. Lett.}\ }\textbf {\bibinfo
  {volume} {68}},\ \bibinfo {pages} {3845} (\bibinfo {year}
  {1992})}\BibitemShut {NoStop}%
\bibitem [{\citenamefont {Archilli}\ \emph {et~al.}(2012)\citenamefont
  {Archilli} \emph {et~al.}}]{Archilli:2011zc}%
  \BibitemOpen
  \bibfield  {author} {\bibinfo {author} {\bibfnamefont {F.}~\bibnamefont
  {Archilli}} \emph {et~al.} (\bibinfo {collaboration} {KLOE-2
  collaboration}),\ }\href {\doibase 10.1016/j.physletb.2011.11.033} {\bibfield
   {journal} {\bibinfo  {journal} {Phys. Lett.}\ }\textbf {\bibinfo {volume}
  {B706}},\ \bibinfo {pages} {251} (\bibinfo {year} {2012})},\ \Eprint
  {http://arxiv.org/abs/1110.0411} {arXiv:1110.0411 [hep-ex]} \BibitemShut
  {NoStop}%
\bibitem [{\citenamefont {Gninenko}(2012{\natexlab{b}})}]{Gninenko:2011uv}%
  \BibitemOpen
  \bibfield  {author} {\bibinfo {author} {\bibfnamefont {S.}~\bibnamefont
  {Gninenko}},\ }\href {\doibase 10.1103/PhysRevD.85.055027} {\bibfield
  {journal} {\bibinfo  {journal} {Phys. Rev.}\ }\textbf {\bibinfo {volume}
  {D85}},\ \bibinfo {pages} {055027} (\bibinfo {year} {2012}{\natexlab{b}})},\
  \Eprint {http://arxiv.org/abs/1112.5438} {arXiv:1112.5438 [hep-ph]}
  \BibitemShut {NoStop}%
\bibitem [{\citenamefont {Babusci}\ \emph {et~al.}(2013)\citenamefont {Babusci}
  \emph {et~al.}}]{Babusci:2012cr}%
  \BibitemOpen
  \bibfield  {author} {\bibinfo {author} {\bibfnamefont {D.}~\bibnamefont
  {Babusci}} \emph {et~al.} (\bibinfo {collaboration} {KLOE-2 collaboration}),\
  }\href {\doibase 10.1016/j.physletb.2013.01.067} {\bibfield  {journal}
  {\bibinfo  {journal} {Phys. Lett.}\ }\textbf {\bibinfo {volume} {B720}},\
  \bibinfo {pages} {111} (\bibinfo {year} {2013})},\ \Eprint
  {http://arxiv.org/abs/1210.3927} {arXiv:1210.3927 [hep-ex]} \BibitemShut
  {NoStop}%
\bibitem [{\citenamefont {Adlarson}\ \emph {et~al.}(2013)\citenamefont
  {Adlarson} \emph {et~al.}}]{Adlarson:2013eza}%
  \BibitemOpen
  \bibfield  {author} {\bibinfo {author} {\bibfnamefont {P.}~\bibnamefont
  {Adlarson}} \emph {et~al.} (\bibinfo {collaboration} {WASA-at-COSY
  collaboration}),\ }\href {\doibase 10.1016/j.physletb.2013.08.055} {\bibfield
   {journal} {\bibinfo  {journal} {Phys. Lett.}\ }\textbf {\bibinfo {volume}
  {B726}},\ \bibinfo {pages} {187} (\bibinfo {year} {2013})},\ \Eprint
  {http://arxiv.org/abs/1304.0671} {arXiv:1304.0671 [hep-ex]} \BibitemShut
  {NoStop}%
\bibitem [{\citenamefont {Agakishiev}\ \emph {et~al.}(2014)\citenamefont
  {Agakishiev} \emph {et~al.}}]{Agakishiev:2013fwl}%
  \BibitemOpen
  \bibfield  {author} {\bibinfo {author} {\bibfnamefont {G.}~\bibnamefont
  {Agakishiev}} \emph {et~al.} (\bibinfo {collaboration} {HADES
  collaboration}),\ }\href {\doibase 10.1016/j.physletb.2014.02.035} {\bibfield
   {journal} {\bibinfo  {journal} {Phys. Lett.}\ }\textbf {\bibinfo {volume}
  {B731}},\ \bibinfo {pages} {265} (\bibinfo {year} {2014})},\ \Eprint
  {http://arxiv.org/abs/1311.0216} {arXiv:1311.0216 [hep-ex]} \BibitemShut
  {NoStop}%
\bibitem [{\citenamefont {Adare}\ \emph {et~al.}(2015)\citenamefont {Adare}
  \emph {et~al.}}]{Adare:2014mgk}%
  \BibitemOpen
  \bibfield  {author} {\bibinfo {author} {\bibfnamefont {A.}~\bibnamefont
  {Adare}} \emph {et~al.} (\bibinfo {collaboration} {PHENIX collaboration}),\
  }\href {\doibase 10.1103/PhysRevC.91.031901} {\bibfield  {journal} {\bibinfo
  {journal} {Phys. Rev.}\ }\textbf {\bibinfo {volume} {C91}},\ \bibinfo {pages}
  {031901} (\bibinfo {year} {2015})},\ \Eprint {http://arxiv.org/abs/1409.0851}
  {arXiv:1409.0851 [nucl-ex]} \BibitemShut {NoStop}%
\bibitem [{\citenamefont {Batley}\ \emph {et~al.}(2015)\citenamefont {Batley}
  \emph {et~al.}}]{Batley:2015lha}%
  \BibitemOpen
  \bibfield  {author} {\bibinfo {author} {\bibfnamefont {J.~R.}\ \bibnamefont
  {Batley}} \emph {et~al.} (\bibinfo {collaboration} {NA48/2}),\ }\href
  {\doibase 10.1016/j.physletb.2015.04.068} {\bibfield  {journal} {\bibinfo
  {journal} {Phys. Lett.}\ }\textbf {\bibinfo {volume} {B746}},\ \bibinfo
  {pages} {178} (\bibinfo {year} {2015})},\ \Eprint
  {http://arxiv.org/abs/1504.00607} {arXiv:1504.00607 [hep-ex]} \BibitemShut
  {NoStop}%
\bibitem [{\citenamefont {Anastasi}\ \emph {et~al.}(2016)\citenamefont
  {Anastasi} \emph {et~al.}}]{KLOE:2016lwm}%
  \BibitemOpen
  \bibfield  {author} {\bibinfo {author} {\bibfnamefont {A.}~\bibnamefont
  {Anastasi}} \emph {et~al.} (\bibinfo {collaboration} {KLOE-2}),\ }\href
  {\doibase 10.1016/j.physletb.2016.04.019} {\bibfield  {journal} {\bibinfo
  {journal} {Phys. Lett.}\ }\textbf {\bibinfo {volume} {B757}},\ \bibinfo
  {pages} {356} (\bibinfo {year} {2016})},\ \Eprint
  {http://arxiv.org/abs/1603.06086} {arXiv:1603.06086 [hep-ex]} \BibitemShut
  {NoStop}%
\bibitem [{\citenamefont {Essig}\ \emph {et~al.}(2011)\citenamefont {Essig},
  \citenamefont {Schuster}, \citenamefont {Toro},\ and\ \citenamefont
  {Wojtsekhowski}}]{Essig:2010xa}%
  \BibitemOpen
  \bibfield  {author} {\bibinfo {author} {\bibfnamefont {R.}~\bibnamefont
  {Essig}}, \bibinfo {author} {\bibfnamefont {P.}~\bibnamefont {Schuster}},
  \bibinfo {author} {\bibfnamefont {N.}~\bibnamefont {Toro}}, \ and\ \bibinfo
  {author} {\bibfnamefont {B.}~\bibnamefont {Wojtsekhowski}},\ }\href {\doibase
  10.1007/JHEP02(2011)009} {\bibfield  {journal} {\bibinfo  {journal} {JHEP}\
  }\textbf {\bibinfo {volume} {02}},\ \bibinfo {pages} {009} (\bibinfo {year}
  {2011})},\ \Eprint {http://arxiv.org/abs/1001.2557} {arXiv:1001.2557
  [hep-ph]} \BibitemShut {NoStop}%
\bibitem [{\citenamefont {Freytsis}\ \emph
  {et~al.}(2010{\natexlab{a}})\citenamefont {Freytsis}, \citenamefont
  {Ovanesyan},\ and\ \citenamefont {Thaler}}]{Freytsis:2009bh}%
  \BibitemOpen
  \bibfield  {author} {\bibinfo {author} {\bibfnamefont {M.}~\bibnamefont
  {Freytsis}}, \bibinfo {author} {\bibfnamefont {G.}~\bibnamefont {Ovanesyan}},
  \ and\ \bibinfo {author} {\bibfnamefont {J.}~\bibnamefont {Thaler}},\ }\href
  {\doibase 10.1007/JHEP01(2010)111} {\bibfield  {journal} {\bibinfo  {journal}
  {JHEP}\ }\textbf {\bibinfo {volume} {01}},\ \bibinfo {pages} {111} (\bibinfo
  {year} {2010}{\natexlab{a}})},\ \Eprint {http://arxiv.org/abs/0909.2862}
  {arXiv:0909.2862 [hep-ph]} \BibitemShut {NoStop}%
\bibitem [{\citenamefont {Balewski}\ \emph {et~al.}(2013)\citenamefont
  {Balewski}, \citenamefont {Bernauer}, \citenamefont {Bertozzi}, \citenamefont
  {Bessuille}, \citenamefont {Buck} \emph {et~al.}}]{Balewski:2013oza}%
  \BibitemOpen
  \bibfield  {author} {\bibinfo {author} {\bibfnamefont {J.}~\bibnamefont
  {Balewski}}, \bibinfo {author} {\bibfnamefont {J.}~\bibnamefont {Bernauer}},
  \bibinfo {author} {\bibfnamefont {W.}~\bibnamefont {Bertozzi}}, \bibinfo
  {author} {\bibfnamefont {J.}~\bibnamefont {Bessuille}}, \bibinfo {author}
  {\bibfnamefont {B.}~\bibnamefont {Buck}},  \emph {et~al.},\ }\href@noop {} {\
   (\bibinfo {year} {2013})},\ \Eprint {http://arxiv.org/abs/1307.4432}
  {arXiv:1307.4432} \BibitemShut {NoStop}%
\bibitem [{\citenamefont {Wojtsekhowski}\ \emph {et~al.}(2012)\citenamefont
  {Wojtsekhowski}, \citenamefont {Nikolenko},\ and\ \citenamefont
  {Rachek}}]{Wojtsekhowski:2012zq}%
  \BibitemOpen
  \bibfield  {author} {\bibinfo {author} {\bibfnamefont {B.}~\bibnamefont
  {Wojtsekhowski}}, \bibinfo {author} {\bibfnamefont {D.}~\bibnamefont
  {Nikolenko}}, \ and\ \bibinfo {author} {\bibfnamefont {I.}~\bibnamefont
  {Rachek}},\ }\href@noop {} {\  (\bibinfo {year} {2012})},\ \Eprint
  {http://arxiv.org/abs/1207.5089} {arXiv:1207.5089 [hep-ex]} \BibitemShut
  {NoStop}%
\bibitem [{\citenamefont {Beranek}\ \emph {et~al.}(2013)\citenamefont
  {Beranek}, \citenamefont {Merkel},\ and\ \citenamefont
  {Vanderhaeghen}}]{Beranek:2013yqa}%
  \BibitemOpen
  \bibfield  {author} {\bibinfo {author} {\bibfnamefont {T.}~\bibnamefont
  {Beranek}}, \bibinfo {author} {\bibfnamefont {H.}~\bibnamefont {Merkel}}, \
  and\ \bibinfo {author} {\bibfnamefont {M.}~\bibnamefont {Vanderhaeghen}},\
  }\href {\doibase 10.1103/PhysRevD.88.015032} {\bibfield  {journal} {\bibinfo
  {journal} {Phys. Rev.}\ }\textbf {\bibinfo {volume} {D88}},\ \bibinfo {pages}
  {015032} (\bibinfo {year} {2013})},\ \Eprint {http://arxiv.org/abs/1303.2540}
  {arXiv:1303.2540 [hep-ph]} \BibitemShut {NoStop}%
\bibitem [{\citenamefont {Echenard}\ \emph {et~al.}(2015)\citenamefont
  {Echenard}, \citenamefont {Essig},\ and\ \citenamefont
  {Zhong}}]{Echenard:2014lma}%
  \BibitemOpen
  \bibfield  {author} {\bibinfo {author} {\bibfnamefont {B.}~\bibnamefont
  {Echenard}}, \bibinfo {author} {\bibfnamefont {R.}~\bibnamefont {Essig}}, \
  and\ \bibinfo {author} {\bibfnamefont {Y.-M.}\ \bibnamefont {Zhong}},\ }\href
  {\doibase 10.1007/JHEP01(2015)113} {\bibfield  {journal} {\bibinfo  {journal}
  {JHEP}\ }\textbf {\bibinfo {volume} {01}},\ \bibinfo {pages} {113} (\bibinfo
  {year} {2015})},\ \Eprint {http://arxiv.org/abs/1411.1770} {arXiv:1411.1770
  [hep-ph]} \BibitemShut {NoStop}%
\bibitem [{\citenamefont {Battaglieri}\ \emph {et~al.}(2015)\citenamefont
  {Battaglieri} \emph {et~al.}}]{Battaglieri:2014hga}%
  \BibitemOpen
  \bibfield  {author} {\bibinfo {author} {\bibfnamefont {M.}~\bibnamefont
  {Battaglieri}} \emph {et~al.},\ }\href {\doibase 10.1016/j.nima.2014.12.017}
  {\bibfield  {journal} {\bibinfo  {journal} {Nucl. Instrum. Meth.}\ }\textbf
  {\bibinfo {volume} {A777}},\ \bibinfo {pages} {91} (\bibinfo {year}
  {2015})},\ \Eprint {http://arxiv.org/abs/1406.6115} {arXiv:1406.6115
  [physics.ins-det]} \BibitemShut {NoStop}%
\bibitem [{\citenamefont {Curtin}\ \emph {et~al.}(2015)\citenamefont {Curtin},
  \citenamefont {Essig}, \citenamefont {Gori},\ and\ \citenamefont
  {Shelton}}]{Curtin:2014cca}%
  \BibitemOpen
  \bibfield  {author} {\bibinfo {author} {\bibfnamefont {D.}~\bibnamefont
  {Curtin}}, \bibinfo {author} {\bibfnamefont {R.}~\bibnamefont {Essig}},
  \bibinfo {author} {\bibfnamefont {S.}~\bibnamefont {Gori}}, \ and\ \bibinfo
  {author} {\bibfnamefont {J.}~\bibnamefont {Shelton}},\ }\href {\doibase
  10.1007/JHEP02(2015)157} {\bibfield  {journal} {\bibinfo  {journal} {JHEP}\
  }\textbf {\bibinfo {volume} {02}},\ \bibinfo {pages} {157} (\bibinfo {year}
  {2015})},\ \Eprint {http://arxiv.org/abs/1412.0018} {arXiv:1412.0018
  [hep-ph]} \BibitemShut {NoStop}%
\bibitem [{\citenamefont {Alekhin}\ \emph {et~al.}(2015)\citenamefont {Alekhin}
  \emph {et~al.}}]{Alekhin:2015byh}%
  \BibitemOpen
  \bibfield  {author} {\bibinfo {author} {\bibfnamefont {S.}~\bibnamefont
  {Alekhin}} \emph {et~al.},\ }\href@noop {} {\  (\bibinfo {year} {2015})},\
  \Eprint {http://arxiv.org/abs/1504.04855} {arXiv:1504.04855 [hep-ph]}
  \BibitemShut {NoStop}%
\bibitem [{\citenamefont {Gardner}\ \emph {et~al.}(2015)\citenamefont
  {Gardner}, \citenamefont {Holt},\ and\ \citenamefont
  {Tadepalli}}]{Gardner:2015wea}%
  \BibitemOpen
  \bibfield  {author} {\bibinfo {author} {\bibfnamefont {S.}~\bibnamefont
  {Gardner}}, \bibinfo {author} {\bibfnamefont {R.~J.}\ \bibnamefont {Holt}}, \
  and\ \bibinfo {author} {\bibfnamefont {A.~S.}\ \bibnamefont {Tadepalli}},\
  }\href@noop {} {\  (\bibinfo {year} {2015})},\ \Eprint
  {http://arxiv.org/abs/1509.00050} {arXiv:1509.00050 [hep-ph]} \BibitemShut
  {NoStop}%
\bibitem [{\citenamefont {Ilten}\ \emph {et~al.}(2015)\citenamefont {Ilten},
  \citenamefont {Thaler}, \citenamefont {Williams},\ and\ \citenamefont
  {Xue}}]{Ilten:2015hya}%
  \BibitemOpen
  \bibfield  {author} {\bibinfo {author} {\bibfnamefont {P.}~\bibnamefont
  {Ilten}}, \bibinfo {author} {\bibfnamefont {J.}~\bibnamefont {Thaler}},
  \bibinfo {author} {\bibfnamefont {M.}~\bibnamefont {Williams}}, \ and\
  \bibinfo {author} {\bibfnamefont {W.}~\bibnamefont {Xue}},\ }\href {\doibase
  10.1103/PhysRevD.92.115017} {\bibfield  {journal} {\bibinfo  {journal} {Phys.
  Rev.}\ }\textbf {\bibinfo {volume} {D92}},\ \bibinfo {pages} {115017}
  (\bibinfo {year} {2015})},\ \Eprint {http://arxiv.org/abs/1509.06765}
  {arXiv:1509.06765 [hep-ph]} \BibitemShut {NoStop}%
\bibitem [{Hea()}]{Hearty}%
  \BibitemOpen
  \href@noop {} {}\bibinfo {note} {Christopher Hearty, private
  communication.}\BibitemShut {Stop}%
\bibitem [{\citenamefont {Benson}\ \emph {et~al.}(2015)\citenamefont {Benson},
  \citenamefont {Gligorov}, \citenamefont {Vesterinen},\ and\ \citenamefont
  {Williams}}]{Benson:2015yzo}%
  \BibitemOpen
  \bibfield  {author} {\bibinfo {author} {\bibfnamefont {S.}~\bibnamefont
  {Benson}}, \bibinfo {author} {\bibfnamefont {V.}~\bibnamefont {Gligorov}},
  \bibinfo {author} {\bibfnamefont {M.~A.}\ \bibnamefont {Vesterinen}}, \ and\
  \bibinfo {author} {\bibfnamefont {M.}~\bibnamefont {Williams}},\ }\bibfield
  {booktitle} {\emph {\bibinfo {booktitle} {{Proceedings, 21st International
  Conference on Computing in High Energy and Nuclear Physics (CHEP 2015)}}},\
  }\href {\doibase 10.1088/1742-6596/664/8/082004} {\bibfield  {journal}
  {\bibinfo  {journal} {J. Phys. Conf. Ser.}\ }\textbf {\bibinfo {volume}
  {664}},\ \bibinfo {pages} {082004} (\bibinfo {year} {2015})}\BibitemShut
  {NoStop}%
\bibitem [{\citenamefont {Aaij}\ \emph {et~al.}(2016)\citenamefont {Aaij} \emph
  {et~al.}}]{Aaij:2015bpa}%
  \BibitemOpen
  \bibfield  {author} {\bibinfo {author} {\bibfnamefont {R.}~\bibnamefont
  {Aaij}} \emph {et~al.} (\bibinfo {collaboration} {LHCb}),\ }\href {\doibase
  10.1007/JHEP03(2016)159} {\bibfield  {journal} {\bibinfo  {journal} {JHEP}\
  }\textbf {\bibinfo {volume} {03}},\ \bibinfo {pages} {159} (\bibinfo {year}
  {2016})},\ \Eprint {http://arxiv.org/abs/1510.01707} {arXiv:1510.01707
  [hep-ex]} \BibitemShut {NoStop}%
\bibitem [{\citenamefont {Barello}\ \emph {et~al.}(2015)\citenamefont
  {Barello}, \citenamefont {Chang},\ and\ \citenamefont
  {Newby}}]{Barello:2015bhq}%
  \BibitemOpen
  \bibfield  {author} {\bibinfo {author} {\bibfnamefont {G.}~\bibnamefont
  {Barello}}, \bibinfo {author} {\bibfnamefont {S.}~\bibnamefont {Chang}}, \
  and\ \bibinfo {author} {\bibfnamefont {C.~A.}\ \bibnamefont {Newby}},\
  }\href@noop {} {\  (\bibinfo {year} {2015})},\ \Eprint
  {http://arxiv.org/abs/1511.02865} {arXiv:1511.02865 [hep-ph]} \BibitemShut
  {NoStop}%
\bibitem [{\citenamefont {Cassel}\ \emph {et~al.}(2010)\citenamefont {Cassel},
  \citenamefont {Ghilencea},\ and\ \citenamefont {Ross}}]{Cassel:2009pu}%
  \BibitemOpen
  \bibfield  {author} {\bibinfo {author} {\bibfnamefont {S.}~\bibnamefont
  {Cassel}}, \bibinfo {author} {\bibfnamefont {D.~M.}\ \bibnamefont
  {Ghilencea}}, \ and\ \bibinfo {author} {\bibfnamefont {G.~G.}\ \bibnamefont
  {Ross}},\ }\href {\doibase 10.1016/j.nuclphysb.2009.10.029} {\bibfield
  {journal} {\bibinfo  {journal} {Nucl. Phys.}\ }\textbf {\bibinfo {volume}
  {B827}},\ \bibinfo {pages} {256} (\bibinfo {year} {2010})},\ \Eprint
  {http://arxiv.org/abs/0903.1118} {arXiv:0903.1118 [hep-ph]} \BibitemShut
  {NoStop}%
\bibitem [{\citenamefont {Cline}\ \emph {et~al.}(2014)\citenamefont {Cline},
  \citenamefont {Dupuis}, \citenamefont {Liu},\ and\ \citenamefont
  {Xue}}]{Cline:2014dwa}%
  \BibitemOpen
  \bibfield  {author} {\bibinfo {author} {\bibfnamefont {J.~M.}\ \bibnamefont
  {Cline}}, \bibinfo {author} {\bibfnamefont {G.}~\bibnamefont {Dupuis}},
  \bibinfo {author} {\bibfnamefont {Z.}~\bibnamefont {Liu}}, \ and\ \bibinfo
  {author} {\bibfnamefont {W.}~\bibnamefont {Xue}},\ }\href {\doibase
  10.1007/JHEP08(2014)131} {\bibfield  {journal} {\bibinfo  {journal} {JHEP}\
  }\textbf {\bibinfo {volume} {08}},\ \bibinfo {pages} {131} (\bibinfo {year}
  {2014})},\ \Eprint {http://arxiv.org/abs/1405.7691} {arXiv:1405.7691
  [hep-ph]} \BibitemShut {NoStop}%
\bibitem [{\citenamefont {Olive}\ \emph {et~al.}(date)\citenamefont {Olive}
  \emph {et~al.}}]{Agashe:2014kda}%
  \BibitemOpen
  \bibfield  {author} {\bibinfo {author} {\bibfnamefont {K.~A.}\ \bibnamefont
  {Olive}} \emph {et~al.} (\bibinfo {collaboration} {Particle Data Group}),\
  }\href {\doibase 10.1088/1674-1137/38/9/090001} {\bibfield  {journal}
  {\bibinfo  {journal} {Chin. Phys.}\ }\textbf {\bibinfo {volume} {C38}},\
  \bibinfo {pages} {090001} (\bibinfo {year} {(2014) and 2015
  update})}\BibitemShut {NoStop}%
\bibitem [{\citenamefont {Abelev}\ \emph {et~al.}(2012)\citenamefont {Abelev}
  \emph {et~al.}}]{ALICE:2011ad}%
  \BibitemOpen
  \bibfield  {author} {\bibinfo {author} {\bibfnamefont {B.}~\bibnamefont
  {Abelev}} \emph {et~al.} (\bibinfo {collaboration} {ALICE}),\ }\href
  {\doibase 10.1016/j.physletb.2012.03.038} {\bibfield  {journal} {\bibinfo
  {journal} {Phys. Lett.}\ }\textbf {\bibinfo {volume} {B710}},\ \bibinfo
  {pages} {557} (\bibinfo {year} {2012})},\ \Eprint
  {http://arxiv.org/abs/1112.2222} {arXiv:1112.2222 [nucl-ex]} \BibitemShut
  {NoStop}%
\bibitem [{SUP()}]{SUP}%
  \BibitemOpen
  \href@noop {} {}\bibinfo {note} {Supplemental material to this
  letter.}\BibitemShut {Stop}%
\bibitem [{LHC(2014)}]{LHCb-TDR-016}%
  \BibitemOpen
  \href@noop {} {\enquote {\bibinfo {title} {{LHCb Trigger and Online Technical
  Design Report}},}\ } (\bibinfo {year} {2014}),\ \bibinfo {note}
  {{LHCb-TDR-016}}\BibitemShut {NoStop}%
\bibitem [{\citenamefont {Alves}\ \emph {et~al.}(2008)\citenamefont {Alves}
  \emph {et~al.}}]{Alves:2008zz}%
  \BibitemOpen
  \bibfield  {author} {\bibinfo {author} {\bibfnamefont {A.~A.}\ \bibnamefont
  {Alves}, \bibfnamefont {Jr.}} \emph {et~al.} (\bibinfo {collaboration}
  {LHCb}),\ }\href {\doibase 10.1088/1748-0221/3/08/S08005} {\bibfield
  {journal} {\bibinfo  {journal} {JINST}\ }\textbf {\bibinfo {volume} {3}},\
  \bibinfo {pages} {S08005} (\bibinfo {year} {2008})}\BibitemShut {NoStop}%
\bibitem [{\citenamefont {Aaij}\ \emph
  {et~al.}(2015{\natexlab{a}})\citenamefont {Aaij} \emph
  {et~al.}}]{Aaij:2014jba}%
  \BibitemOpen
  \bibfield  {author} {\bibinfo {author} {\bibfnamefont {R.}~\bibnamefont
  {Aaij}} \emph {et~al.} (\bibinfo {collaboration} {LHCb}),\ }\href {\doibase
  10.1142/S0217751X15300227} {\bibfield  {journal} {\bibinfo  {journal} {Int.
  J. Mod. Phys.}\ }\textbf {\bibinfo {volume} {A30}},\ \bibinfo {pages}
  {1530022} (\bibinfo {year} {2015}{\natexlab{a}})},\ \Eprint
  {http://arxiv.org/abs/1412.6352} {arXiv:1412.6352 [hep-ex]} \BibitemShut
  {NoStop}%
\bibitem [{\citenamefont {Aaij}\ \emph {et~al.}(2013)\citenamefont {Aaij} \emph
  {et~al.}}]{Aaij:2012rt}%
  \BibitemOpen
  \bibfield  {author} {\bibinfo {author} {\bibfnamefont {R.}~\bibnamefont
  {Aaij}} \emph {et~al.} (\bibinfo {collaboration} {LHCb}),\ }\href {\doibase
  10.1007/JHEP01(2013)090} {\bibfield  {journal} {\bibinfo  {journal} {JHEP}\
  }\textbf {\bibinfo {volume} {01}},\ \bibinfo {pages} {090} (\bibinfo {year}
  {2013})},\ \Eprint {http://arxiv.org/abs/1209.4029} {arXiv:1209.4029
  [hep-ex]} \BibitemShut {NoStop}%
\bibitem [{\citenamefont {Aaij}\ \emph
  {et~al.}(2015{\natexlab{b}})\citenamefont {Aaij} \emph
  {et~al.}}]{Aaij:2014azz}%
  \BibitemOpen
  \bibfield  {author} {\bibinfo {author} {\bibfnamefont {R.}~\bibnamefont
  {Aaij}} \emph {et~al.} (\bibinfo {collaboration} {LHCb}),\ }\href {\doibase
  10.1007/JHEP02(2015)121} {\bibfield  {journal} {\bibinfo  {journal} {JHEP}\
  }\textbf {\bibinfo {volume} {02}},\ \bibinfo {pages} {121} (\bibinfo {year}
  {2015}{\natexlab{b}})},\ \Eprint {http://arxiv.org/abs/1409.8548}
  {arXiv:1409.8548 [hep-ex]} \BibitemShut {NoStop}%
\bibitem [{\citenamefont {Cacciari}\ \emph {et~al.}(2008)\citenamefont
  {Cacciari}, \citenamefont {Salam},\ and\ \citenamefont
  {Soyez}}]{Cacciari:2008gp}%
  \BibitemOpen
  \bibfield  {author} {\bibinfo {author} {\bibfnamefont {M.}~\bibnamefont
  {Cacciari}}, \bibinfo {author} {\bibfnamefont {G.~P.}\ \bibnamefont {Salam}},
  \ and\ \bibinfo {author} {\bibfnamefont {G.}~\bibnamefont {Soyez}},\ }\href
  {\doibase 10.1088/1126-6708/2008/04/063} {\bibfield  {journal} {\bibinfo
  {journal} {JHEP}\ }\textbf {\bibinfo {volume} {04}},\ \bibinfo {pages} {063}
  (\bibinfo {year} {2008})},\ \Eprint {http://arxiv.org/abs/0802.1189}
  {arXiv:0802.1189 [hep-ph]} \BibitemShut {NoStop}%
\bibitem [{\citenamefont {Cacciari}\ \emph {et~al.}(2012)\citenamefont
  {Cacciari}, \citenamefont {Salam},\ and\ \citenamefont
  {Soyez}}]{Cacciari:2011ma}%
  \BibitemOpen
  \bibfield  {author} {\bibinfo {author} {\bibfnamefont {M.}~\bibnamefont
  {Cacciari}}, \bibinfo {author} {\bibfnamefont {G.~P.}\ \bibnamefont {Salam}},
  \ and\ \bibinfo {author} {\bibfnamefont {G.}~\bibnamefont {Soyez}},\ }\href
  {\doibase 10.1140/epjc/s10052-012-1896-2} {\bibfield  {journal} {\bibinfo
  {journal} {Eur. Phys. J.}\ }\textbf {\bibinfo {volume} {C72}},\ \bibinfo
  {pages} {1896} (\bibinfo {year} {2012})},\ \Eprint
  {http://arxiv.org/abs/1111.6097} {arXiv:1111.6097 [hep-ph]} \BibitemShut
  {NoStop}%
\bibitem [{LHC(2013)}]{LHCb-TDR-013}%
  \BibitemOpen
  \href@noop {} {\enquote {\bibinfo {title} {{LHCb VELO Upgrade Technical
  Design Report}},}\ } (\bibinfo {year} {2013}),\ \bibinfo {note}
  {{LHCb-TDR-013}}\BibitemShut {NoStop}%
\bibitem [{\citenamefont {Sjöstrand}\ \emph {et~al.}(2015)\citenamefont
  {Sjöstrand}, \citenamefont {Ask}, \citenamefont {Christiansen},
  \citenamefont {Corke}, \citenamefont {Desai}, \citenamefont {Ilten},
  \citenamefont {Mrenna}, \citenamefont {Prestel}, \citenamefont {Rasmussen},\
  and\ \citenamefont {Skands}}]{Sjostrand:2014zea}%
  \BibitemOpen
  \bibfield  {author} {\bibinfo {author} {\bibfnamefont {T.}~\bibnamefont
  {Sjöstrand}}, \bibinfo {author} {\bibfnamefont {S.}~\bibnamefont {Ask}},
  \bibinfo {author} {\bibfnamefont {J.~R.}\ \bibnamefont {Christiansen}},
  \bibinfo {author} {\bibfnamefont {R.}~\bibnamefont {Corke}}, \bibinfo
  {author} {\bibfnamefont {N.}~\bibnamefont {Desai}}, \bibinfo {author}
  {\bibfnamefont {P.}~\bibnamefont {Ilten}}, \bibinfo {author} {\bibfnamefont
  {S.}~\bibnamefont {Mrenna}}, \bibinfo {author} {\bibfnamefont
  {S.}~\bibnamefont {Prestel}}, \bibinfo {author} {\bibfnamefont {C.~O.}\
  \bibnamefont {Rasmussen}}, \ and\ \bibinfo {author} {\bibfnamefont {P.~Z.}\
  \bibnamefont {Skands}},\ }\href {\doibase 10.1016/j.cpc.2015.01.024}
  {\bibfield  {journal} {\bibinfo  {journal} {Comput. Phys. Commun.}\ }\textbf
  {\bibinfo {volume} {191}},\ \bibinfo {pages} {159} (\bibinfo {year}
  {2015})},\ \Eprint {http://arxiv.org/abs/1410.3012} {arXiv:1410.3012
  [hep-ph]} \BibitemShut {NoStop}%
\bibitem [{\citenamefont {Aad}\ \emph {et~al.}(2014)\citenamefont {Aad} \emph
  {et~al.}}]{Aad:2014zya}%
  \BibitemOpen
  \bibfield  {author} {\bibinfo {author} {\bibfnamefont {G.}~\bibnamefont
  {Aad}} \emph {et~al.} (\bibinfo {collaboration} {ATLAS}),\ }\href {\doibase
  10.1140/epjc/s10052-014-3034-9} {\bibfield  {journal} {\bibinfo  {journal}
  {Eur. Phys. J.}\ }\textbf {\bibinfo {volume} {C74}},\ \bibinfo {pages} {3034}
  (\bibinfo {year} {2014})},\ \Eprint {http://arxiv.org/abs/1404.4562}
  {arXiv:1404.4562 [hep-ex]} \BibitemShut {NoStop}%
\bibitem [{\citenamefont {Chatrchyan}\ \emph {et~al.}(2012)\citenamefont
  {Chatrchyan} \emph {et~al.}}]{Chatrchyan:2012xi}%
  \BibitemOpen
  \bibfield  {author} {\bibinfo {author} {\bibfnamefont {S.}~\bibnamefont
  {Chatrchyan}} \emph {et~al.} (\bibinfo {collaboration} {CMS}),\ }\href
  {\doibase 10.1088/1748-0221/7/10/P10002} {\bibfield  {journal} {\bibinfo
  {journal} {JINST}\ }\textbf {\bibinfo {volume} {7}},\ \bibinfo {pages}
  {P10002} (\bibinfo {year} {2012})},\ \Eprint {http://arxiv.org/abs/1206.4071}
  {arXiv:1206.4071 [physics.ins-det]} \BibitemShut {NoStop}%
\bibitem [{\citenamefont {Aaij}\ \emph
  {et~al.}(2011{\natexlab{a}})\citenamefont {Aaij} \emph
  {et~al.}}]{Aaij:2011uk}%
  \BibitemOpen
  \bibfield  {author} {\bibinfo {author} {\bibfnamefont {R.}~\bibnamefont
  {Aaij}} \emph {et~al.} (\bibinfo {collaboration} {LHCb}),\ }\href {\doibase
  10.1016/j.physletb.2011.08.017} {\bibfield  {journal} {\bibinfo  {journal}
  {Phys. Lett.}\ }\textbf {\bibinfo {volume} {B703}},\ \bibinfo {pages} {267}
  (\bibinfo {year} {2011}{\natexlab{a}})},\ \Eprint
  {http://arxiv.org/abs/1107.3935} {arXiv:1107.3935 [hep-ex]} \BibitemShut
  {NoStop}%
\bibitem [{\citenamefont {Aaij}\ \emph
  {et~al.}(2011{\natexlab{b}})\citenamefont {Aaij} \emph
  {et~al.}}]{Aaij:2011jh}%
  \BibitemOpen
  \bibfield  {author} {\bibinfo {author} {\bibfnamefont {R.}~\bibnamefont
  {Aaij}} \emph {et~al.} (\bibinfo {collaboration} {LHCb}),\ }\href {\doibase
  10.1140/epjc/s10052-011-1645-y} {\bibfield  {journal} {\bibinfo  {journal}
  {Eur. Phys. J.}\ }\textbf {\bibinfo {volume} {C71}},\ \bibinfo {pages} {1645}
  (\bibinfo {year} {2011}{\natexlab{b}})},\ \Eprint
  {http://arxiv.org/abs/1103.0423} {arXiv:1103.0423 [hep-ex]} \BibitemShut
  {NoStop}%
\bibitem [{\citenamefont {Aaij}\ \emph {et~al.}(2012)\citenamefont {Aaij} \emph
  {et~al.}}]{LHCb:2012aa}%
  \BibitemOpen
  \bibfield  {author} {\bibinfo {author} {\bibfnamefont {R.}~\bibnamefont
  {Aaij}} \emph {et~al.} (\bibinfo {collaboration} {LHCb}),\ }\href {\doibase
  10.1140/epjc/s10052-012-2025-y} {\bibfield  {journal} {\bibinfo  {journal}
  {Eur. Phys. J.}\ }\textbf {\bibinfo {volume} {C72}},\ \bibinfo {pages} {2025}
  (\bibinfo {year} {2012})},\ \Eprint {http://arxiv.org/abs/1202.6579}
  {arXiv:1202.6579 [hep-ex]} \BibitemShut {NoStop}%
\bibitem [{\citenamefont {LHCb}(2012)}]{LHCb:2012fja}%
  \BibitemOpen
  \bibfield  {author} {\bibinfo {author} {\bibnamefont {LHCb}},\ }\href@noop {}
  {\bibfield  {journal} {\bibinfo  {journal} {LHCb-CONF-2012-013,
  CERN-LHCb-CONF-2012-013}\ } (\bibinfo {year} {2012})}\BibitemShut {NoStop}%
\bibitem [{\citenamefont {Williams}(2015)}]{Williams:2015xfa}%
  \BibitemOpen
  \bibfield  {author} {\bibinfo {author} {\bibfnamefont {M.}~\bibnamefont
  {Williams}},\ }\href {\doibase 10.1088/1748-0221/10/06/P06002} {\bibfield
  {journal} {\bibinfo  {journal} {JINST}\ }\textbf {\bibinfo {volume} {10}},\
  \bibinfo {pages} {P06002} (\bibinfo {year} {2015})},\ \Eprint
  {http://arxiv.org/abs/1503.04767} {arXiv:1503.04767 [hep-ex]} \BibitemShut
  {NoStop}%
\bibitem [{\citenamefont {Freytsis}\ \emph
  {et~al.}(2010{\natexlab{b}})\citenamefont {Freytsis}, \citenamefont
  {Ligeti},\ and\ \citenamefont {Thaler}}]{Freytsis:2009ct}%
  \BibitemOpen
  \bibfield  {author} {\bibinfo {author} {\bibfnamefont {M.}~\bibnamefont
  {Freytsis}}, \bibinfo {author} {\bibfnamefont {Z.}~\bibnamefont {Ligeti}}, \
  and\ \bibinfo {author} {\bibfnamefont {J.}~\bibnamefont {Thaler}},\ }\href
  {\doibase 10.1103/PhysRevD.81.034001} {\bibfield  {journal} {\bibinfo
  {journal} {Phys. Rev.}\ }\textbf {\bibinfo {volume} {D81}},\ \bibinfo {pages}
  {034001} (\bibinfo {year} {2010}{\natexlab{b}})},\ \Eprint
  {http://arxiv.org/abs/0911.5355} {arXiv:0911.5355 [hep-ph]} \BibitemShut
  {NoStop}%
\bibitem [{\citenamefont {Aaij}\ \emph
  {et~al.}(2015{\natexlab{c}})\citenamefont {Aaij} \emph
  {et~al.}}]{Aaij:2015tna}%
  \BibitemOpen
  \bibfield  {author} {\bibinfo {author} {\bibfnamefont {R.}~\bibnamefont
  {Aaij}} \emph {et~al.} (\bibinfo {collaboration} {LHCb}),\ }\href {\doibase
  10.1103/PhysRevLett.115.161802} {\bibfield  {journal} {\bibinfo  {journal}
  {Phys. Rev. Lett.}\ }\textbf {\bibinfo {volume} {115}},\ \bibinfo {pages}
  {161802} (\bibinfo {year} {2015}{\natexlab{c}})},\ \Eprint
  {http://arxiv.org/abs/1508.04094} {arXiv:1508.04094 [hep-ex]} \BibitemShut
  {NoStop}%
\bibitem [{\citenamefont {Haisch}\ and\ \citenamefont
  {Kamenik}(2016)}]{Haisch:2016hzu}%
  \BibitemOpen
  \bibfield  {author} {\bibinfo {author} {\bibfnamefont {U.}~\bibnamefont
  {Haisch}}\ and\ \bibinfo {author} {\bibfnamefont {J.~F.}\ \bibnamefont
  {Kamenik}},\ }\href {\doibase 10.1103/PhysRevD.93.055047} {\bibfield
  {journal} {\bibinfo  {journal} {Phys. Rev.}\ }\textbf {\bibinfo {volume}
  {D93}},\ \bibinfo {pages} {055047} (\bibinfo {year} {2016})},\ \Eprint
  {http://arxiv.org/abs/1601.05110} {arXiv:1601.05110 [hep-ph]} \BibitemShut
  {NoStop}%
\bibitem [{\citenamefont {Adinolfi}\ \emph {et~al.}(2013)\citenamefont
  {Adinolfi} \emph {et~al.}}]{LHCb-DP-2012-003}%
  \BibitemOpen
  \bibfield  {author} {\bibinfo {author} {\bibfnamefont {M.}~\bibnamefont
  {Adinolfi}} \emph {et~al.},\ }\href {\doibase 10.1140/epjc/s10052-013-2431-9}
  {\bibfield  {journal} {\bibinfo  {journal} {Eur. Phys. J.}\ }\textbf
  {\bibinfo {volume} {C73}},\ \bibinfo {pages} {2431} (\bibinfo {year}
  {2013})},\ \Eprint {http://arxiv.org/abs/1211.6759} {arXiv:1211.6759
  [physics.ins-det]} \BibitemShut {NoStop}%
\bibitem [{\citenamefont {Dent}\ \emph {et~al.}(2012)\citenamefont {Dent},
  \citenamefont {Ferrer},\ and\ \citenamefont {Krauss}}]{Dent:2012mx}%
  \BibitemOpen
  \bibfield  {author} {\bibinfo {author} {\bibfnamefont {J.~B.}\ \bibnamefont
  {Dent}}, \bibinfo {author} {\bibfnamefont {F.}~\bibnamefont {Ferrer}}, \ and\
  \bibinfo {author} {\bibfnamefont {L.~M.}\ \bibnamefont {Krauss}},\
  }\href@noop {} {\  (\bibinfo {year} {2012})},\ \Eprint
  {http://arxiv.org/abs/1201.2683} {arXiv:1201.2683 [astro-ph.CO]} \BibitemShut
  {NoStop}%
\bibitem [{\citenamefont {Kazanas}\ \emph {et~al.}(2014)\citenamefont
  {Kazanas}, \citenamefont {Mohapatra}, \citenamefont {Nussinov}, \citenamefont
  {Teplitz},\ and\ \citenamefont {Zhang}}]{Kazanas:2014mca}%
  \BibitemOpen
  \bibfield  {author} {\bibinfo {author} {\bibfnamefont {D.}~\bibnamefont
  {Kazanas}}, \bibinfo {author} {\bibfnamefont {R.~N.}\ \bibnamefont
  {Mohapatra}}, \bibinfo {author} {\bibfnamefont {S.}~\bibnamefont {Nussinov}},
  \bibinfo {author} {\bibfnamefont {V.~L.}\ \bibnamefont {Teplitz}}, \ and\
  \bibinfo {author} {\bibfnamefont {Y.}~\bibnamefont {Zhang}},\ }\href
  {\doibase 10.1016/j.nuclphysb.2014.11.009} {\bibfield  {journal} {\bibinfo
  {journal} {Nucl. Phys.}\ }\textbf {\bibinfo {volume} {B890}},\ \bibinfo
  {pages} {17} (\bibinfo {year} {2014})},\ \Eprint
  {http://arxiv.org/abs/1410.0221} {arXiv:1410.0221 [hep-ph]} \BibitemShut
  {NoStop}%
\end{thebibliography}%

\section*{Supplemental Material}

Here, we provide a more detailed discussion on the following aspects of our proposed $A'$ search at LHCb:  how to subtract the misidentified di-muon background; how to estimate the LHCb muon-identification performance from \Ref{Aaij:2014azz}; what the consistent-decay-topology requirements are in our analysis; how the sensitivity of LHCb to $A'\to\mu^+\mu^-$ could be improved; and the reach plotted with an extended $\epsilon^2$ range. 

\subsection{Fake Di-Muon Subtraction}

Most fake muons come from misidentified pions, with a subdominant contribution from misidentified kaons and protons.  For simplicity, we denote all misID particles as pions below, though the argument presented is completely general.  The following logic will allow us to use data-driven methods to subtract the fake di-muon background.

We first consider the double-misID case where two pions are each misidentified as muons.  The number of same-sign (SS) $\pi^{\pm}\pi^{\pm}$ pairs from a $pp$ collision is related to the number of pions that satisfy our kinematic requirements  by
\be
n_{\pm\pm}^{\pi\pi} = \frac{n_{\pm}^{\pi}(n_{\pm}^{\pi}-1)}{2},
\ee
while the number of opposite-sign (OS) $\pi^{+}\pi^{-}$ pairs is
\be
n_{+-}^{\pi\pi} = n_+^{\pi}n_-^{\pi}.
\ee
In the $n_{\pm}^{\pi}\to\infty$ limit, and assuming equal acceptance for SS and OS pairs with the same invariant mass, we obtain the simple relationship 
\be
n_{+-}^{\pi\pi} \approx 2 \sqrt{n_{++}^{\pi\pi} n_{--}^{\pi\pi}} \approx n_{++}^{\pi\pi} +  n_{--}^{\pi\pi},
\ee
where the right-most relationship assumes $n_+^\pi \approx n_-^{\pi}$, which is a good approximation in $pp$ collisions. Therefore, the total number of SS pion pairs in a data sample is
\be
N_{+-}^{\pi\pi} \approx N_{++}^{\pi\pi} +  N_{--}^{\pi\pi},
\ee
where $N_{xy} \equiv \sum n_{xy}$ and the sum is over all collisions in the sample.  

Next, we consider the single-misID case where one true muon is combined with a misidentified prompt pion.   This true muon dominantly comes from a displaced heavy-flavor decay which is mis-reconstructed as prompt.  In this case, the combinatorics only enter for the pion, and in the full data sample we find
\be
N_{+-}^{\pi\mu} \approx N_{++}^{\pi\mu} +  N_{--}^{\pi\mu}.
\ee

Combining the double- and single-misID cases together, one expects
\be
\label{eq:appssos}
B^{\pi\pi}_{\rm misID} + B^{\pi\mu}_{\rm misID} \approx N_{++} +  N_{--},
\ee
where the lack of superscripts on $N_{\pm\pm}$ denotes that we do not need to separate these into $\pi\pi$ and $\pi\mu$ categories experimentally.  This simple estimate, based on taking the asymptotic limit and assuming charge-symmetric pion samples, could easily be improved in an actual analysis, since the true combinatorics can be determined from the data.  The small correction required to account for the difference in acceptance between SS and OS pairs can be obtained reliably from simulation. We expect that \eqref{eq:appssos} is accurate to $\approx 10\%$ and that a highly-accurate misID subtraction can be performed using the data.   

Finally, we note that an alternative approach is also possible using OS $\pi\pi$ samples directly with the pion misID rate measured in data, along with OS $\mu\pi$ samples where the muon is displaced.  The actual analysis could use both methods and check their consistency to assess the systematic uncertainty in the fake-muon background subtraction.

\subsection{Muon Identification}

For small $m_{A'}$, most $A' \to \mu^+\mu^-$ decays produce low-$p_T$ muons. Since high-energy $pp$ collisions produce many low-$p_T$ pions, there are many possible $\pi^+\pi^-$ pairs per collision that could result in a double misID of $\pi^+\pi^-$ as $\mu^+\mu^-$. 
Furthermore, the decay-in-flight probability of $\pi \to \mu$ is inversely proportional to momentum.
Therefore, the low-mass $A' \to \mu^+\mu^-$ signal is obscured by the enormous double-misID $\pi^+\pi^-$ background if muon identification is based soley on whether the particle is a muon when it reaches the muon system.  
Our baseline selection requires $p_T(\mu) > 0.5\gev$, $p(\mu) > 10\gev$, and $\eta(\mu) \in [2,5]$.  
By convolving the pion momentum spectrum obtained from {\sc Pythia} with the decay-in-flight probability given by the pion lifetime, we obtain an estimate that $\approx 1\%$ of all pions satisfying these kinematic requirements will be identified as muons by the muon system.
This results in $B^{\pi\pi}_{\rm misID} = \mathcal{O}(100)\times B_{\rm EM}$ in the low-$m_{A'}$ region. 

One way to reduce the double-misID background is to increase the muon $p_T$ threshold.   At low-$m_{A'}$, however, the signal is predominantly produced via $\eta \to A' \gamma$, so increasing the muon $p_T$ threshold greatly reduces the potential signal yield.
For example, increasing the muon $p_T$ threshold from 0.5\gev, as in our nominal proposed search, to 2\gev reduces the low-$m_{A'}$ yield by a factor of $\approx 100$.  That said, such an approach may prove viable at ATLAS and CMS as they plan to collect 200 times more luminosity by the end of the HL-LHC era than LHCb will collect in Run~3, making it plausible that the low-$m_{A'}$ reach estimate in this letter could be representative of the ultimate ATLAS/CMS sensitivity.

Instead of raising the  muon $p_T$ threshold, here we take advantage of the unique particle-identification features of LHCb.  The LHCb detector employs two ring-imaging Cherenkov (RICH) detectors to identify charged particles with momenta from $\mathcal{O}(1-100\gev)$. 
The primary motivation for incorporating such detector systems into LHCb was to provide hadronic particle-identification capabilities to facilitate studying Cabbibo-suppressed weak decays.  For our purposes, these systems are also very powerful tools for lepton identification.  For $p \lesssim 5\gev$, the RICH detectors are capable of identifying electrons and muons without the need for additional information from the calorimeter or muon systems.  
By combining the information of the RICH detectors with all other LHCb subsystems into a neural network (NN), LHCb is able to greatly reduce the pion (and kaon) misID probabilities~\cite{Aaij:2014azz}.

To our knowledge, the only published performance of the LHCb muon-identification NN is from a search for the decay $\tau \to 3\mu$~\cite{Aaij:2014azz}.
The muon kinematic requirements in that analysis are similar to ours, and so we estimate the NN performance directly from \Ref{Aaij:2014azz}. 
Specifically, we assume the 2012 performance and a requirement that removes the lowest two bins in NN response (see Fig.~2d of \Ref{Aaij:2014azz}).
The efficiency to identify a true di-muon pair is taken to be $\varepsilon(\tau \to 3\mu)^{2/3} \approx 54\%$, which we reduce to 50\% to account for other selection criteria applied to $A' \to \mu^+\mu^-$ candidates.
Since the $\tau \to 3\mu$ selection only used displaced tracks, the background sample dominantly contains at least one true muon.  Assuming that all background candidates are $\mu\pi\pi$ gives $\varepsilon(\pi\pi) \approx 10^{-6}$; this value includes the probability of decay-in-flight $\pi\to\mu$ of $\approx 1\%$ per pion.  Since the background likely contains a non-negligible fraction of $\mu\mu\pi$ candidates, this is an underestimate of the single-pion rejection from the NN.  Therefore, we obtain a conservative estimate of the reach in $m_{A'}$ regions where $B_{\rm misID}$ is important. 
Finally, we note that using a simple likelihood-based approach, as in \Ref{LHCb-DP-2012-003}, LHCb obtains a per-pion misID rate that is only a factor of two higher for the same muon efficiency than the NN value used in our analysis.  

\subsection{Decay-Topology Criteria}

One of the key ingredients in our analysis is to enforce a consistent $A'$ decay topology.  These requirements are the same as in \Ref{Ilten:2015hya}, but with electrons replaced by muons; we repeat them here for the convenience of the reader.  
We also apply one additional criterion here that is not used in \Ref{Ilten:2015hya}: displaced di-muon vertices are rejected if a third displaced particle within LHCb acceptance and satisfying $p_T > 0.5\gev$ has a distance of closest approach (DOCA) less than 1~mm relative to either muon; this reduces the heavy-flavor background. 

The one-dimensional track IP resolution expected in Run~3 is well approximated by~\cite{LHCb-TDR-013}
\be
\sigma_{\rm IP} = \left(11.0 + \frac{13.1\,{\rm GeV}}{p_{\rm T}} \right)\,\mu{\rm m},
\ee
while the resolution on $\ell_{\rm T}$ is 
\begin{equation}
\sigma_{\ell_{\rm T}} \approx \frac{\sin{\theta_{A'}}}{\alpha_{\mu\mu}} \sqrt{\sigma_{\mu^+{\rm IP}}^2 + \sigma_{\mu^-{\rm IP}}^2},
\end{equation}
where $\alpha_{\mu\mu}$ is the $A'$ decay opening angle and the $A'$ is constrained to originate from the $pp$ collision.  As stated in the main text, we require  ${\rm IP}_{\mu^\pm} < 2.5\,\sigma_{{\rm IP}}$ in the prompt $A'$ search.

For the displaced $A'$ searches, we require the reconstructed $A' \to \mu^+ \mu^-$ candidate to satisfy the following consistency requirements:
\begin{itemize}
\item the $A'$ decay vertex is downstream of the $pp$ collision;
\item the DOCA between the two muon tracks is consistent with zero;
\item the angle between $\vec{p}_{A'}$ and the spatial vector formed from the $pp$ collision to the $A'$ decay vertex is consistent with zero;
\item the IP out of the $A'$ decay plane for each muon is consistent with zero, where the decay plane is defined by the $pp$ collision point and the first hits on the $\mu^+$ and $\mu^-$ tracks. 
\end{itemize}
In each case, we define ``consistent with zero'' as having a $p$-value greater than 1\%.  Therefore, the efficiency on a true displaced $A'$ decay is close to 100\%.  
We also apply the DOCA requirement in the prompt search.  

\subsection{Possible Improvements}

\begin{figure*}[!t]
\includegraphics[width=1.7\columnwidth]{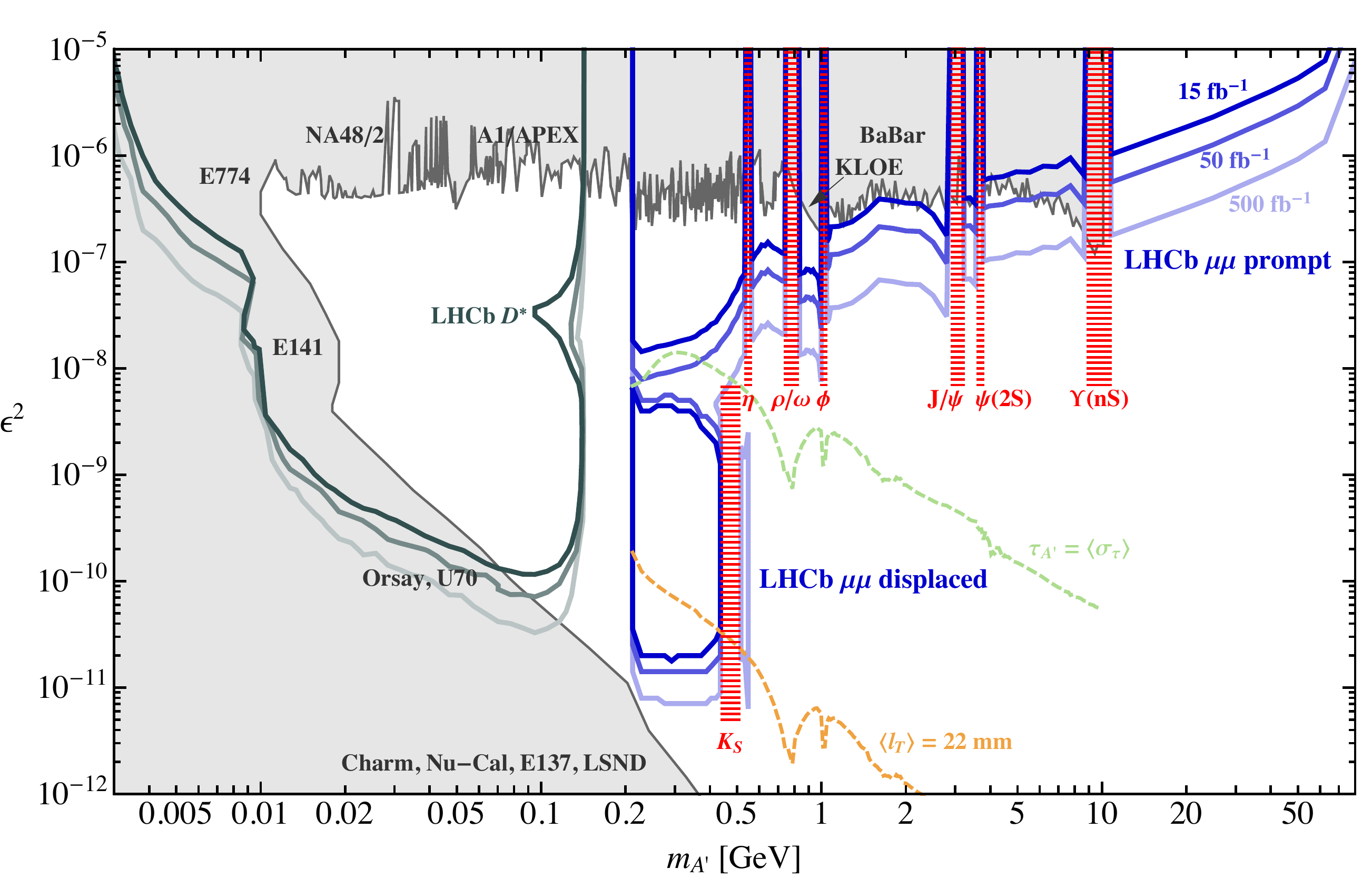}
\caption{Same as Fig.~1 of the main text, but scaling the $15~\text{fb}^{-1}$ baseline up to $50~\text{fb}^{-1}$ and $500~\text{fb}^{-1}$, for both the $D^* \to D^0 A'$ search \cite{Ilten:2015hya} and the inclusive di-muon search (this work).  For reference, the green dashed line shows where the $A'$ lifetime is the same as the LHCb di-muon lifetime resolution and the orange dashed line shows where the average $A'$ transverse displacement matches the distance at which the muons from $A'\to\mu^+\mu^-$ decays likely no longer have enough hits in the LHCb VELO.
}
\label{fig:suppreachmumu}
\end{figure*}

The following improvements could result in an increased sensitivity at LHCb to $A' \to \ell^+\ell^-$ decays:
\begin{itemize}
\item {\it Excited mesons}:  When estimating $B_M$, \textsc{Pythia} does not produce known heavy mesons (e.g.~$\rho(1450)$) through Lund string fragmentation (though they can be produced through heavy-flavor decays).  These excited mesons have been observed to decay to di-electrons, so one would assume that they also decay to di-muons with the same rate up to phase space effects.  The fact that such mesons are not included in our study likely means that we underestimate the reach for $m_{A'} \in [1,3] \gev$.  In fact, it is plausible that such mesons provide the dominant source of potential $A'$ production in this mass region.  If so, then one would likely want to shift the isolation criteria to apply only for $m_{A'} > m_{J/\psi}$ (instead of $m_{A'} > m_{\phi}$ as in the text).
\item {\it Event selection}: In this study, we used a simple and robust cut-based selection strategy. A multivariate approach would likely perform better, especially in the displaced searches where correlations between various features used in the consistent-decay-topology requirements could be exploited. 
Furthermore, the $p_T(\mu) > 0.5\gev$ and $p(\mu) > 10\gev$ requirements discard about 70\% of the $A' \to \mu^+\mu^-$ signal decays that could be fully reconstructed in LHCb.  If the low-$p_T$ misID background can be suppressed in the real-time data-analysis system, then the $A'$ yield could be increased by up to a factor of 3 for the same luminosity.  
\item {\it Search strategy}:  Here, we considered the reach assuming three distinct search regions:  prompt, pre-module, post-module.  One could optimally combine these regions following \Ref{Williams:2015xfa} which should improve the reach in the low-mass region. 
\item {\it Semi-inclusive search}: Instead of using the inclusive di-muon spectrum, a similar search could be done in semi-inclusive hadron decays such as $M\to \ell^+\ell^-Y$, more in the spirit of \Ref{Ilten:2015hya}.  Depending on the channel, one could use the invariant mass of the $M$ or $Y$ system as a constraint to help control fake muon backgrounds.
\item {\it Di-electron search.}  To cover the mass range $m_{A'} \in [2m_e,2m_\mu]$, one could pursue a similar inclusive search strategy for the di-electron final state.  That said, the di-electron mass resolution is significantly degraded by Bremsstrahlung radiation and multiple scattering~\cite{Ilten:2015hya}.  In \Ref{Ilten:2015hya}, the $m_{ee}$ resolution could be improved by imposing the kinematic constraints from charm meson decays, which is not an option in an inclusive search.  For the displaced $A'$ search, these same effects degrade the vertex resolution, and $e^+e^-$ pairs from photon conversion are a challenging background in the post-module region.  For these reasons, we suspect that $A' \to e^+ e^-$ is best probed using an exclusive (or semi-inclusive) strategy, but it would be worth testing the fully inclusive approach on LHCb data.
\item {\it Luminosity}: Our study is based on 15~fb$^{-1}$ of data collected by LHCb, which is a conservative estimate of what is expected in Run~3.  LHCb expects to collect at least 50~fb$^{-1}$ of data in Runs 3 and 4 combined, and may eventually collect 10--30 times more data than considered in this study.  The impact on the dark photon reach from scaling up the LHCb luminosity is shown in \Fig{fig:suppreachmumu}.
\end{itemize}

\subsection{Extended Reach Plot}

To better show the array of proposed dark photon experiments, in \Fig{fig:suppreachmumu} we show the same reach plot from the main text, but with an extended $\epsilon^2$ range including supernova bounds~(SN)~\cite{Dent:2012mx,Kazanas:2014mca}.  

\begin{figure*}[!t]
\includegraphics[width=1.7\columnwidth]{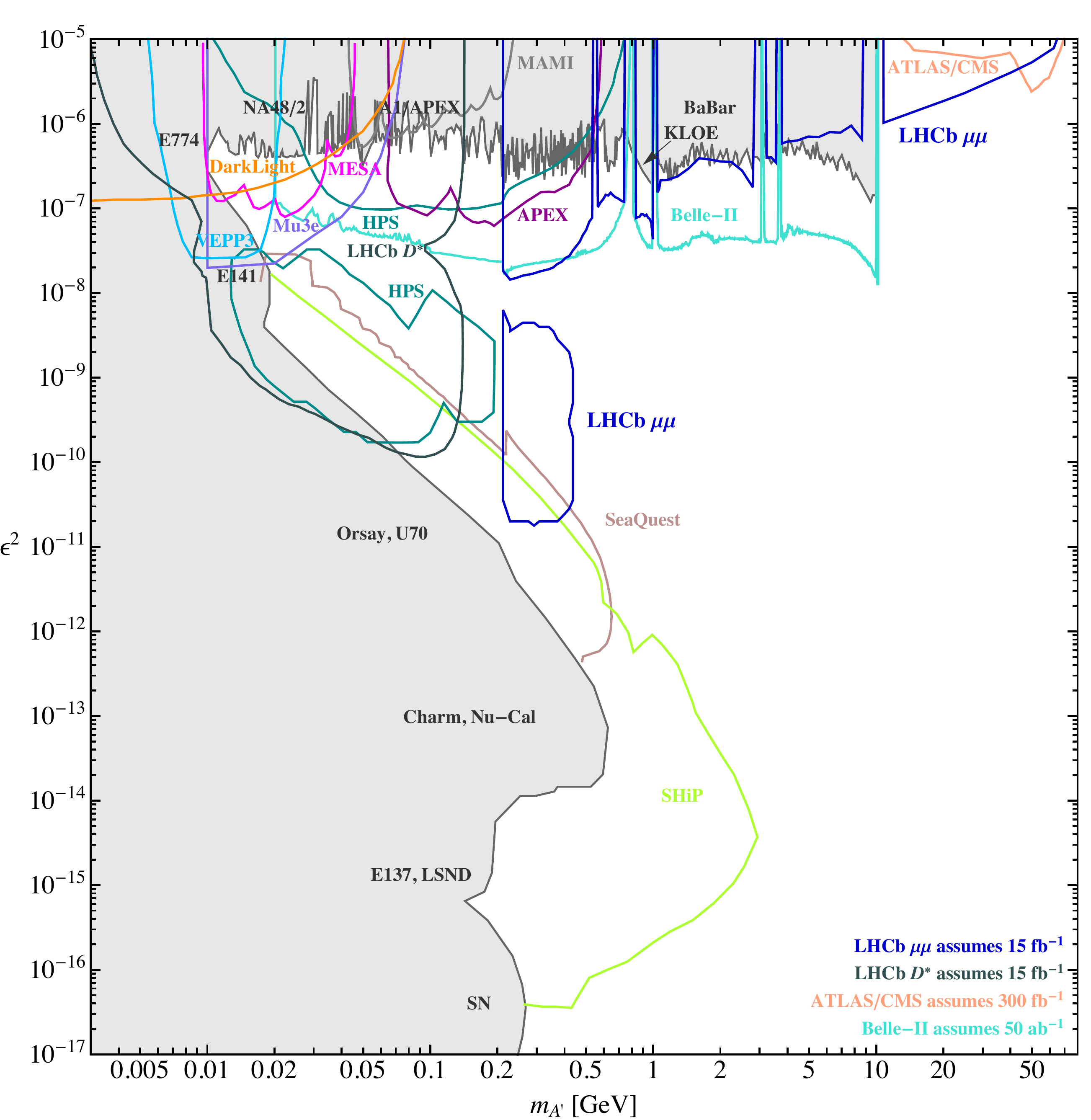}
\caption{Same as Fig.~1 of the main text, but with an extended $\epsilon^2$ range and without the exclusion regions due to narrow resonances. 
}
\label{fig:suppreachmumu}
\end{figure*}

\end{document}